\documentclass[sigconf]{acmart}

\AtBeginDocument{%
  }

\copyrightyear{2026}
\acmYear{2026}
\setcopyright{cc}
\setcctype{by}
\acmConference[CIKM '26]{Proceedings of the 35th ACM International Conference on Information and Knowledge Management}{November 07--11, 2026}{Rome, Italy}
\acmBooktitle{Proceedings of the 35th ACM International Conference on Information and Knowledge Management (CIKM '26), November 07--11, 2026, Rome, Italy}
\acmDOI{10.1145/3799682.3841104}
\acmISBN{979-8-4007-2539-5/2026/11}
\usepackage{multirow}
\usepackage{subcaption}
\newcommand{\cuecell}[4]{\begin{tabular}[t]{@{}p{0.47\linewidth}@{\hspace{0.03\linewidth}}p{0.47\linewidth}@{}}$\bullet$ #1 & $\bullet$ #2\\[-1pt]$\bullet$ #3 & $\bullet$ #4\end{tabular}}

\graphicspath{{./figures/}{figures/}}
\providecommand{\NewFigRoot}{./figures/}
\providecommand{\NewFig}[2][]{\includegraphics[#1]{\NewFigRoot#2}}

\begin{document}

\title[RouteRec for Sequential Recommendation]{RouteRec: Behavior-Guided Sparse Routing for Sequential Recommendation}

\author{Junyeong Song}
\orcid{0009-0009-7590-4673}
\affiliation{%
  \institution{Korea Advanced Institute of Science and Technology}
  \city{Daejeon}
  \country{Republic of Korea}}
\email{jy1559@kaist.ac.kr}

\author{Jaemin Yoo}
\orcid{0000-0001-7237-5117}
\affiliation{%
  \institution{Seoul National University}
  \city{Seoul}
  \country{Republic of Korea}}
\email{jaeminyoo@snu.ac.kr}

\renewcommand{\shortauthors}{Junyeong Song and Jaemin Yoo}

\begin{abstract}
Sessionized interaction histories contain behavioral patterns that can improve sequential recommendation.
However, existing models process all sessions through the same parameterized blocks, regardless of their behavioral differences.
Mixture of Experts (MoE) enables conditional computation, but it leaves open what should guide expert allocation.
We propose RouteRec, a sequential recommender that uses observed session behavior as the routing criterion.
RouteRec summarizes four types of behavioral evidence from sessionized histories: interaction tempo, item-group focus, repetition and carryover, and popularity tendency. It uses these cues to route computation at macro, mid, and micro scopes. Cue-derived scores first select expert groups; within each selected group, the current backbone state then refines expert selection.
Across six public datasets and 18 dataset--metric combinations, RouteRec ranks first in 12 and second in three, yielding the best overall average rank of 1.61 compared with 4.11 for the next-best baseline.
Additional analyses suggest that the behavioral cues guide expert allocation beyond added capacity and produce routing patterns aligned with observed behavior.
Our code is available at \url{https://github.com/jy1559/RouteRec}.
\end{abstract}



\maketitle

\section{Introduction}
\label{sec:intro}

Sequential recommendation predicts the next item from a user's interaction history. Session-based methods use the active session as a short-term sequence, whereas session-aware methods can also use the same user's past sessions.
Previous work has modeled such histories using short-range transition models \cite{rendle2010fpmc}, recurrent sequence models \cite{hidasi2016gru4rec}, self-attentive backbones \cite{kang2018sasrec}, time-aware variants \cite{li2020tisasrec}, and contrastive learning methods \cite{qiu2022duorec}; see Wang et al.~\cite{wang2019seqsurvey} for a survey.

\begin{figure}[!t]
  \centering
  \NewFig[width=\columnwidth]{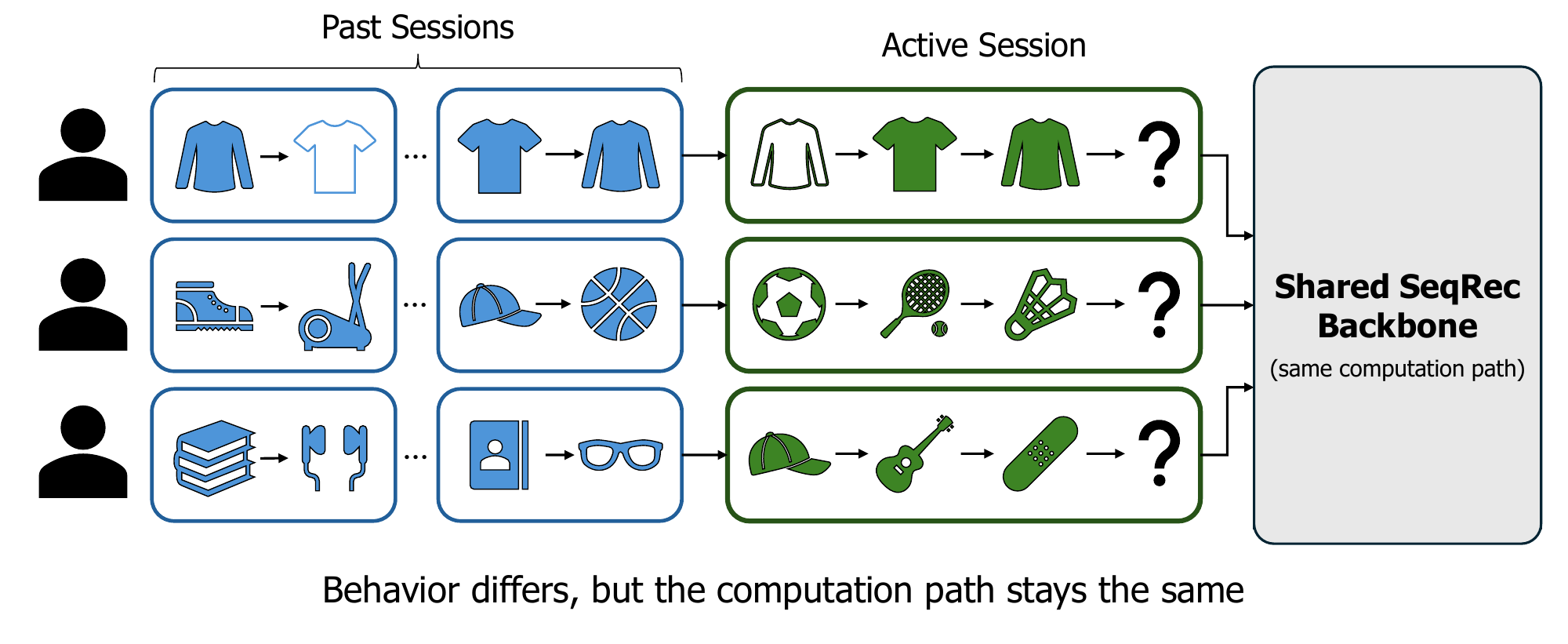}
  \vspace{2pt}
  \NewFig[width=\columnwidth]{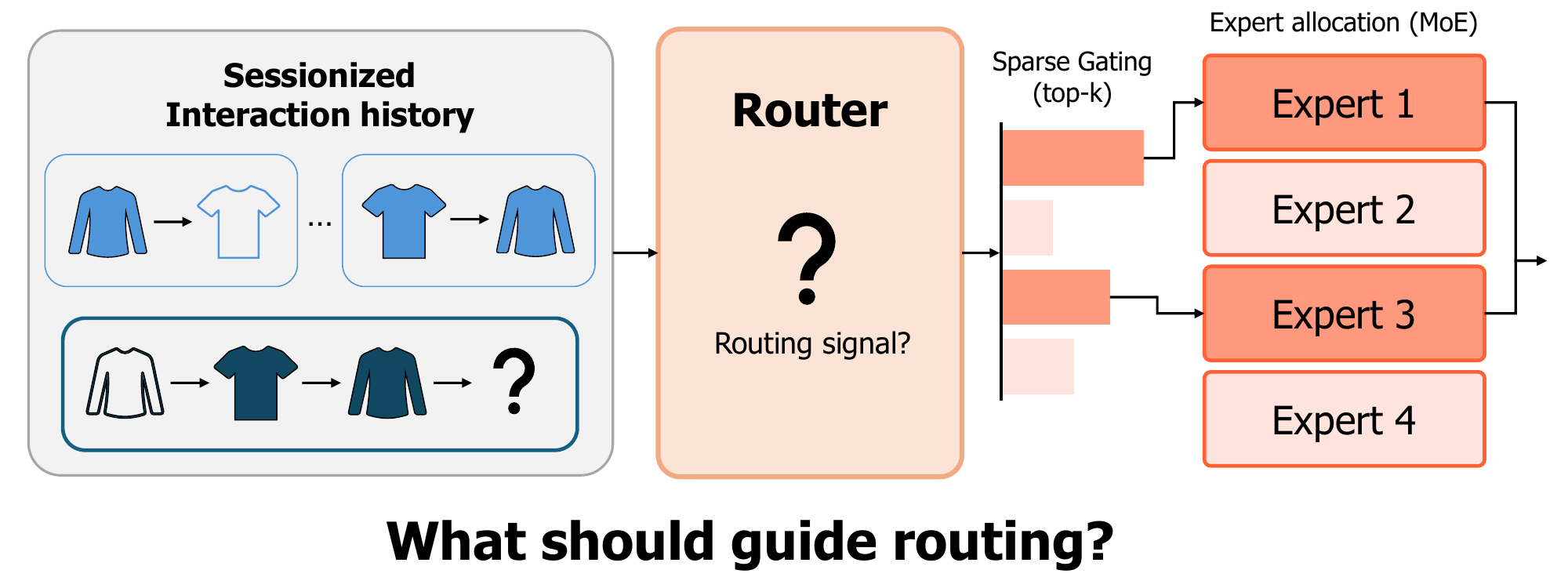}
  \caption{Sessions with different behavioral patterns share one computation path in standard sequential recommendation (\textit{top}), motivating behavior-guided expert allocation via MoE routing (\textit{bottom}).}
  \Description{Two-part motivation figure. The first panel shows sessions with different past and active-session behavior entering the same shared sequential recommendation backbone. The second panel shows a router and sparse expert allocation, asking what information from sessionized history should be used as the router input.}
  \label{fig:intro-behavioral-features}
\end{figure}

Different sessions make different evidence useful for prediction: rapid transitions emphasize recent item changes, whereas repetitive sessions emphasize repeated items and cross-session carryover.
Yet existing models apply the same shared blocks to every session and cannot directly select computation for the active behavior pattern.

Conditional computation allows input-dependent paths. Mixture of Experts (MoE) routes each input to a small subset of specialized experts~\cite{jacobs1991adaptive,shazeer2017moe,fedus2022switch}, so different sessions can follow different paths.

The central question is then how to allocate sessions to experts. Prior MoE work shows that performance depends not only on having experts but also on how inputs are assigned to them~\cite{fedus2022switch}. Existing routers often use predefined criteria such as task labels~\cite{ma2018mmoe} or dataset source~\cite{jain2023damex}, but session-aware recommendation provides no such label. The routing signal must therefore be inferred from the observed sessionized history (Figure~\ref{fig:intro-behavioral-features}).

\begin{figure}[!t]
  \centering
  \begin{subfigure}[t]{0.48\columnwidth}
    \centering
    \NewFig[width=\linewidth]{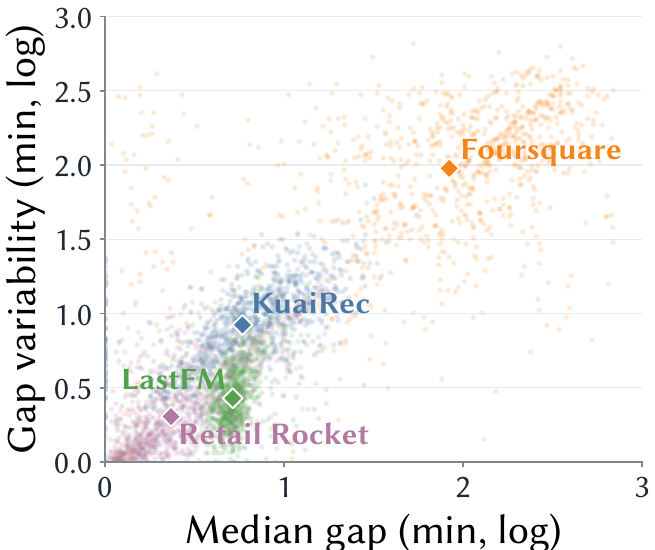}
    \caption{Interaction tempo}
    \label{fig:dataset-stat-evidence-tempo}
  \end{subfigure}%
  \hfill
  \begin{subfigure}[t]{0.48\columnwidth}
    \centering
    \NewFig[width=\linewidth]{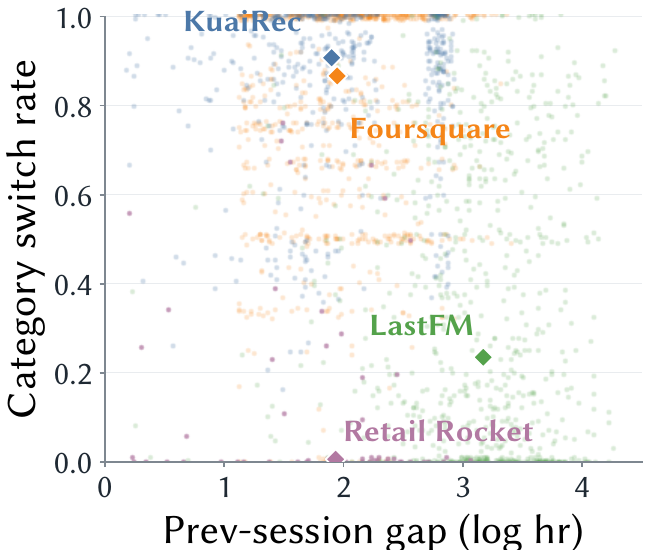}
    \caption{Cross-session recency and movement}
    \label{fig:dataset-stat-evidence-carryover}
  \end{subfigure}
  \caption{Sessions differ along cue-derived behavioral patterns within and across four representative datasets. Each point is a session; diamonds mark dataset medians. This variation motivates using behavioral cues as the routing criterion.}
  \Description{Two subfigures: (a) interaction tempo (median gap vs. gap variability), (b) cross-session recency and movement (prev-session gap vs. category switch rate). Colored point clouds represent KuaiRec, Foursquare, LastFM, and Retail Rocket.}
  \label{fig:dataset-stat-evidence}
\end{figure}

We propose using observed session behavior as the routing signal. It must be available before prediction, derivable from standard logs and metadata without routing labels, and informative about sessions that may benefit from different computation paths.
Unlike a fixed task or domain label, this signal is session-specific: different sessions from the same user can activate different computation paths as their observed behavior changes.
We construct four compact cue families from logs and available metadata: Tempo from timestamps, Focus from item-group changes, Memory from within- and cross-session repetition, and Popularity from training-set frequency. These cues summarize mechanisms studied in prior work on time-aware dynamics, session intent, repeat consumption, and popularity effects~\cite{li2020tisasrec,li2017narm,liu2018stamp,ren2019repeatnet,klimashevskaia2024popularitybias}. Their patterns vary within and across datasets (Figure~\ref{fig:dataset-stat-evidence}).
The cues select expert groups, and the current hidden state refines expert selection within each group. This hierarchical sparse routing operates at macro, mid, and micro scopes, combining behavioral family evidence with position-specific backbone context.

Across six public datasets and nine baselines, RouteRec achieves the best overall average rank and leads all three metrics on KuaiRec, LastFM, and Retail Rocket.

Our main contributions are as follows:
\begin{itemize}
    \item We propose observed session behavior as a routing criterion for expert allocation and represent it with cue families derived from sessionized interaction histories.
    \item We implement this criterion in RouteRec, which uses cue-derived group routing and hidden-conditioned expert selection across macro, mid, and micro scopes.
    \item Across six datasets, RouteRec achieves the best overall average rank over nine baselines. Matched controls separate behavior-guided routing from added capacity and show that behavior-guided routing outperforms hidden-state routing on both SASRec and DuoRec. Expert usage also shifts with the behavioral cues.
\end{itemize}

\section{Problem Definition and Related Work}
\label{sec:related}

\subsection{Problem Definition}
We study session-aware next-item prediction.
For each user, interactions are divided into chronological sessions, and one session is treated as the active session at prediction time.
Given the observed prefix in an active session and the user's earlier sessions, the task is to predict the next item in the active session.

All inputs used by RouteRec are computed only from information available before the prediction target is observed. Item popularity is estimated from the training split. Item-group cues use dataset-provided coarse labels such as category, genre, or artist when available; otherwise, the corresponding cues are zero-filled. No cue uses the held-out target item.

\begin{figure*}[!t]
  \centering
  \NewFig[width=0.95\textwidth]{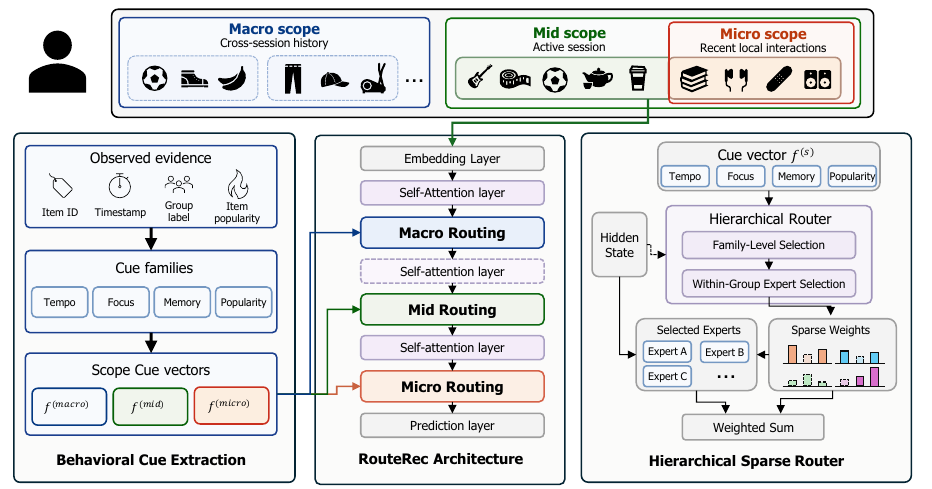}
  \caption{An overview of RouteRec. \textit{Left}: interaction histories become behavioral cue vectors at macro, mid, and micro scopes. \textit{Center}: the SASRec backbone applies routed expert blocks at these scopes. \textit{Right}: family-specific cues select expert groups; the corresponding cues and current hidden state select experts within chosen groups. The hidden state is used only in this within-group stage.}
  \Description{A three-panel architecture figure of RouteRec. The left panel converts observed evidence into cue families and scope cue vectors. The center panel shows an active-session SASRec backbone with macro, mid, and micro routing points. The right panel shows cue-based expert-group selection followed by cue-and-hidden-conditioned expert selection and sparse weighting.}
  \label{fig:framework}
\end{figure*}

\subsection{Sequential, Session-Based, and Session-Aware Recommendation}

Sequential recommendation predicts future items from ordered user histories. Early methods capture local transitions with Markov-chain, recurrent, or convolutional architectures~\cite{rendle2010fpmc,hidasi2016gru4rec,tang2018caser}. Self-attentive models later became common backbones~\cite{kang2018sasrec,sun2019bert4rec}, with recent extensions using contrastive objectives, frequency-aware encoding, or state-space layers~\cite{qiu2022duorec,du2023fearec,liu2025sigma}.

Session-based recommendation focuses on short interaction windows and models current-session intent using attention or graph structures~\cite{li2017narm,liu2018stamp,wu2019srgnn}.
Session-aware recommendation additionally uses earlier sessions of the same user, making the active session and past sessions distinct sources of evidence~\cite{latifi2021sessionaware}.
Related work also enriches sequence representations with side information: FDSA and DIF-SR incorporate item attributes~\cite{zhang2019fdsa,xie2022difsr}, while TiSASRec encodes interaction intervals~\cite{li2020tisasrec}.

RouteRec builds on these advances in sequence representation but asks a different question: should observed session behavior also determine the computation path?

\subsection{MoE in Recommendation}

MoE-based recommendation models use expert decomposition to model heterogeneous signals. M3SRec uses expert mixtures for multimodal fusion~\cite{bian2023m3srec}, while FAME places experts within attention heads to represent facet-specific preferences~\cite{liu2025fame}. HM4SR combines interactive and temporal MoE modules for multimodal sequential recommendation~\cite{zhang2025hm4sr}, and HM2Rec replaces Mamba feed-forward layers with MoE modules~\cite{liu2026hm2rec}. FamouSRec routes behavior sequences through heterogeneous sequential encoders using frequency components of the sequence~\cite{zhang2025famousrec}. Together, these studies establish the value of expert decomposition in recommendation, but the evidence used to route an input remains a separate design choice.

\subsection{MoE Routing Design}

A central question in MoE routing design is what information the router uses to choose expert paths. Standard MoE layers often route from the current hidden representation~\cite{fedus2022switch}. Routers can also use structured signals: multi-task recommenders use task-specific gates~\cite{ma2018mmoe,tang2020ple}, language-aware MoE routes using language identity~\cite{wang2023lrmoe}, and dataset-aware MoE routes by data source~\cite{jain2023damex}. These designs illustrate how a routing signal can correspond to a meaningful difference among inputs.

Session-aware recommendation does not provide a predefined task, dataset, or language label for each session. A routing criterion must therefore come from the interaction history itself. FamouSRec~\cite{zhang2025famousrec} routes among heterogeneous sequential encoders using frequency components of behavior sequences. RouteRec instead represents four behavioral dimensions of sessionized histories---tempo, item-group focus, memory and carryover, and popularity tendency---at macro, mid, and micro scopes.

\section{Proposed Method}
\label{sec:method}

We propose RouteRec, an MoE sequential recommender that uses observed session behavior to guide expert allocation while retaining a standard sequential backbone.

\begin{table*}[t]
    \centering
    \caption{Example cue scalars showing what each behavioral family measures at each routing scope. The observation window differs across scopes. A cue-bank summary is in Appendix~\ref{app:cue-bank}.}
    \label{tab:cue-summary}
    \small
    \setlength{\tabcolsep}{7pt}
    \begin{tabular}{l|ccc}
        \toprule
        \textbf{Family} & \textbf{Macro examples} & \textbf{Mid examples} & \textbf{Micro examples} \\
        \midrule
        Tempo   & session gaps, pace trend         & interval mean/std               & last gap, local pace shift \\
        Focus   & theme consistency                & category concentration, switching & local switch, suffix focus \\
        Memory  & repetition, carryover            & repeat rate, novelty            & recent reuse, longest run \\
        Popularity                       & popularity level, drift         & mean popularity, spread, trend & last-item pop., suffix spread, pop. delta \\
        \bottomrule
    \end{tabular}
\end{table*}
\subsection{Main Ideas and Architecture}
\label{sec:routerec-overview}

RouteRec makes three design choices: behavioral cues, observation windows, and sparse expert selection. We call each window a routing scope and its summary a cue vector. Sections~\ref{sec:feature-construction}--\ref{sec:hierarchical-routing} develop these choices; Table~\ref{tab:cue-summary} gives representative cues.

\paragraph{Overall architecture.}

RouteRec uses SASRec~\cite{kang2018sasrec}, which alternates self-attention and feed-forward blocks, as its backbone. Self-attention captures item dependencies, while feed-forward blocks transform the resulting hidden states. RouteRec retains self-attention and replaces the feed-forward stages at the macro, mid, and micro scopes (Section~\ref{sec:multi-stage-routing}) with routed expert blocks (Figure~\ref{fig:framework}).

At routing scope $s$, let $\ell$ denote its backbone layer and $\mathbf{H}^{(\ell-1)}$ the hidden states entering that layer. Multi-head self-attention and layer normalization produce
\[
\widetilde{\mathbf{H}}^{(\ell)} = \mathrm{AttnBlock}\!\left(\mathbf{H}^{(\ell-1)}\right).
\]

For session $i$ and position $t$, $\mathrm{Router}^{(s)}$ combines cue vector $\mathbf{f}^{(s)}_{i,t}$ with backbone state $\widetilde{\mathbf{h}}^{(\ell)}_{i,t}$ to produce sparse group and expert weights:
\[
\boldsymbol{\pi}^{(s)}_{i,t}
= \mathrm{Router}^{(s)}\!\left(\widetilde{\mathbf{h}}^{(\ell)}_{i,t},\mathbf{f}^{(s)}_{i,t}\right).
\]
Macro and mid cues are shared across positions, while micro cues and hidden-conditioned expert scores are position-specific. The routed block combines selected expert outputs with the residual connection:
\[
\mathbf{h}^{(\ell)}_{i,t}
= \mathrm{LN}\!\left(
\widetilde{\mathbf{h}}^{(\ell)}_{i,t}
+ \sum_{g,k} \pi^{(s)}_{i,t,g,k}\,
\mathrm{Expert}^{(s)}_{g,k}\!\left(\widetilde{\mathbf{h}}^{(\ell)}_{i,t}\right)
\right),
\]
where $\pi^{(s)}_{i,t,g,k}$ weights expert $k$ in group $g$; non-selected experts have zero weight. Section~\ref{sec:hierarchical-routing} details the router.

\subsection{Behavioral Cue Construction}
\label{sec:feature-construction}

RouteRec organizes behavioral patterns observable in sessionized logs into four cue families (Table~\ref{tab:cue-summary}).

\textit{Tempo} captures the pace of interactions and interval patterns from timestamps~\cite{li2020tisasrec}.
\textit{Focus} captures item-group concentration and switching from category, artist, or genre labels, motivated by work on session intent~\cite{li2017narm,liu2018stamp,wu2019srgnn} and attribute-based item grouping~\cite{xie2022difsr,zhang2019fdsa}.
\textit{Memory} captures within-session repetition and cross-session carryover from item identities, motivated by short-term session memory and repeat consumption~\cite{liu2018stamp,ren2019repeatnet}.
\textit{Popularity} captures head-tail tendency from training-set item frequency, a central signal in studies of popularity bias~\cite{klimashevskaia2024popularitybias}.
Each family appears at every scope but summarizes a different window. All cues use only history observed before the target; Appendix~\ref{app:cue-bank} lists the complete bank.

\subsection{Coarse-to-Fine Multi-Scope Routing}
\label{sec:multi-stage-routing}

Session behavior unfolds at multiple time scales. Cross-session history reveals long-run tendencies, the active session shows how a pattern develops before prediction, and the most recent interactions reflect immediate local state. RouteRec therefore assigns one routing scope to each observation window. The \textit{macro} scope summarizes earlier sessions, the \textit{mid} scope the observed active-session prefix, and the \textit{micro} scope the last $w=5$ interactions.

For routing scope $s$, session $i$, and prediction position $t$, the corresponding observed interaction window $\mathcal{R}^{(s)}_{i,t}$ is aggregated into a cue vector:
\[
\mathbf{f}^{(s)}_{i,t} = \mathcal{A}\!\left(\mathcal{R}^{(s)}_{i,t}\right),
\]
where $\mathcal{A}$ is the cue aggregation function. Macro and mid cues are computed once per session and shared across positions; micro cues are re-evaluated at each position.
This separates stable cross-session and active-session context from local changes that may emerge near the prediction point.

Routing scopes are interleaved with self-attention blocks from coarse to fine (Figure~\ref{fig:framework}), letting later scopes refine a representation conditioned on broader evidence. The intervening depth is tuned (Table~\ref{tab:appendix-search-space}); in the selected configurations, the dashed block in Figure~\ref{fig:framework} is usually empty.

\begin{table*}[!t]
  \caption{Next-item prediction performance on six datasets against nine baseline methods. Each entry reports mean $\pm$ sample standard deviation over seeds 42, 43, and 44. Hyperparameters and checkpoints are selected using validation performance, and the test set is evaluated only after selection. \textbf{Bold} and \underline{underline} denote best and second-best competition ranks. Avg.~Rank is computed over all 18 dataset--metric cells (lower is better).}
  \label{tab:main-results}
  \centering
  
  \small
  \setlength{\tabcolsep}{3.0pt}
  \renewcommand{\arraystretch}{1.02}
  \resizebox{\textwidth}{!}{%
  \begin{tabular}{ll*{10}{c}}
    \toprule
    Dataset & Metric & \multicolumn{1}{c}{SASRec} & \multicolumn{1}{c}{GRU4Rec} & \multicolumn{1}{c}{TiSASRec} & \multicolumn{1}{c}{FEARec} & \multicolumn{1}{c}{DuoRec} & \multicolumn{1}{c}{BSARec} & \multicolumn{1}{c}{FAME} & \multicolumn{1}{c}{DIF-SR} & \multicolumn{1}{c}{FDSA} & \multicolumn{1}{c}{\textbf{RouteRec}} \\
    \midrule
    \multirow{3}{*}{KuaiRec} & HR@10 & .4612{\scriptsize$\pm$}.0028 & .4352{\scriptsize$\pm$}.0108 & .3299{\scriptsize$\pm$}.0201 & .3653{\scriptsize$\pm$}.0309 & .4779{\scriptsize$\pm$}.0107 & .4174{\scriptsize$\pm$}.0098 & \underline{.4809}{\scriptsize$\pm$}.0017 & .3201{\scriptsize$\pm$}.0655 & .4671{\scriptsize$\pm$}.0092 & \textbf{.4836}{\scriptsize$\pm$}.0035 \\
     & NDCG@10 & .4319{\scriptsize$\pm$}.0028 & .4020{\scriptsize$\pm$}.0115 & .2682{\scriptsize$\pm$}.0200 & .3101{\scriptsize$\pm$}.0419 & .4555{\scriptsize$\pm$}.0168 & .3739{\scriptsize$\pm$}.0155 & \underline{.4617}{\scriptsize$\pm$}.0015 & .2802{\scriptsize$\pm$}.0887 & .4377{\scriptsize$\pm$}.0137 & \textbf{.4656}{\scriptsize$\pm$}.0041 \\
     & MRR@10 & .4229{\scriptsize$\pm$}.0029 & .3918{\scriptsize$\pm$}.0117 & .2491{\scriptsize$\pm$}.0199 & .2932{\scriptsize$\pm$}.0452 & .4486{\scriptsize$\pm$}.0187 & .3604{\scriptsize$\pm$}.0171 & \underline{.4558}{\scriptsize$\pm$}.0016 & .2679{\scriptsize$\pm$}.0959 & .4286{\scriptsize$\pm$}.0151 & \textbf{.4602}{\scriptsize$\pm$}.0043 \\
    \midrule
    \multirow{3}{*}{LastFM} & HR@10 & .3439{\scriptsize$\pm$}.0018 & .3281{\scriptsize$\pm$}.0017 & .3353{\scriptsize$\pm$}.0011 & .2978{\scriptsize$\pm$}.0022 & \underline{.3447}{\scriptsize$\pm$}.0010 & .2278{\scriptsize$\pm$}.0013 & .2465{\scriptsize$\pm$}.0011 & .2884{\scriptsize$\pm$}.0014 & .3321{\scriptsize$\pm$}.0014 & \textbf{.3569}{\scriptsize$\pm$}.0006 \\
     & NDCG@10 & \underline{.2703}{\scriptsize$\pm$}.0018 & .2645{\scriptsize$\pm$}.0019 & .2644{\scriptsize$\pm$}.0005 & .2260{\scriptsize$\pm$}.0017 & .2577{\scriptsize$\pm$}.0010 & .1572{\scriptsize$\pm$}.0018 & .1846{\scriptsize$\pm$}.0020 & .2114{\scriptsize$\pm$}.0014 & .2639{\scriptsize$\pm$}.0009 & \textbf{.2866}{\scriptsize$\pm$}.0012 \\
     & MRR@10 & \underline{.2474}{\scriptsize$\pm$}.0018 & .2448{\scriptsize$\pm$}.0020 & .2425{\scriptsize$\pm$}.0004 & .2038{\scriptsize$\pm$}.0015 & .2306{\scriptsize$\pm$}.0010 & .1353{\scriptsize$\pm$}.0021 & .1654{\scriptsize$\pm$}.0023 & .1875{\scriptsize$\pm$}.0015 & .2427{\scriptsize$\pm$}.0008 & \textbf{.2646}{\scriptsize$\pm$}.0015 \\
    \midrule    
     \multirow{3}{*}{Retail Rocket} & HR@10 & \underline{.5440}{\scriptsize$\pm$}.0027 & .4931{\scriptsize$\pm$}.0028 & .5350{\scriptsize$\pm$}.0019 & .5371{\scriptsize$\pm$}.0023 & .5235{\scriptsize$\pm$}.0045 & .4536{\scriptsize$\pm$}.0031 & .4365{\scriptsize$\pm$}.0054 & .5318{\scriptsize$\pm$}.0012 & .5306{\scriptsize$\pm$}.0021 & \textbf{.5462}{\scriptsize$\pm$}.0032 \\
     & NDCG@10 & .3730{\scriptsize$\pm$}.0014 & .3329{\scriptsize$\pm$}.0024 & .3649{\scriptsize$\pm$}.0007 & \underline{.3776}{\scriptsize$\pm$}.0045 & .3597{\scriptsize$\pm$}.0030 & .3525{\scriptsize$\pm$}.0033 & .3521{\scriptsize$\pm$}.0012 & .3730{\scriptsize$\pm$}.0034 & .3622{\scriptsize$\pm$}.0027 & \textbf{.3805}{\scriptsize$\pm$}.0022 \\
     & MRR@10 & .3194{\scriptsize$\pm$}.0016 & .2832{\scriptsize$\pm$}.0022 & .3116{\scriptsize$\pm$}.0012 & \underline{.3274}{\scriptsize$\pm$}.0056 & .3085{\scriptsize$\pm$}.0029 & .3201{\scriptsize$\pm$}.0033 & .3247{\scriptsize$\pm$}.0006 & .3231{\scriptsize$\pm$}.0047 & .3095{\scriptsize$\pm$}.0028 & \textbf{.3285}{\scriptsize$\pm$}.0019 \\
    \midrule
    \multirow{3}{*}{ML-1M} & HR@10 & .1572{\scriptsize$\pm$}.0063 & .1567{\scriptsize$\pm$}.0033 & \textbf{.1616}{\scriptsize$\pm$}.0019 & .1389{\scriptsize$\pm$}.0087 & .1490{\scriptsize$\pm$}.0032 & .1557{\scriptsize$\pm$}.0077 & .1522{\scriptsize$\pm$}.0065 & .1575{\scriptsize$\pm$}.0002 & .1581{\scriptsize$\pm$}.0055 & \underline{.1585}{\scriptsize$\pm$}.0033 \\
     & NDCG@10 & .0781{\scriptsize$\pm$}.0035 & .0839{\scriptsize$\pm$}.0024 & .0835{\scriptsize$\pm$}.0019 & .0682{\scriptsize$\pm$}.0024 & .0756{\scriptsize$\pm$}.0021 & .0837{\scriptsize$\pm$}.0021 & .0818{\scriptsize$\pm$}.0054 & .0830{\scriptsize$\pm$}.0007 & \underline{.0840}{\scriptsize$\pm$}.0020 & \textbf{.0846}{\scriptsize$\pm$}.0022 \\
     & MRR@10 & .0542{\scriptsize$\pm$}.0027 & .0618{\scriptsize$\pm$}.0026 & .0601{\scriptsize$\pm$}.0028 & .0467{\scriptsize$\pm$}.0015 & .0535{\scriptsize$\pm$}.0021 & \underline{.0619}{\scriptsize$\pm$}.0007 & .0606{\scriptsize$\pm$}.0050 & .0606{\scriptsize$\pm$}.0005 & .0616{\scriptsize$\pm$}.0011 & \textbf{.0624}{\scriptsize$\pm$}.0022 \\
    \midrule
    \multirow{3}{*}{Foursquare} & HR@10 & .6207{\scriptsize$\pm$}.0152 & .5527{\scriptsize$\pm$}.0168 & .6332{\scriptsize$\pm$}.0208 & .6054{\scriptsize$\pm$}.0133 & .6456{\scriptsize$\pm$}.0129 & .4952{\scriptsize$\pm$}.0185 & .4914{\scriptsize$\pm$}.0224 & .4770{\scriptsize$\pm$}.0201 & \underline{.6514}{\scriptsize$\pm$}.0066 & \textbf{.6533}{\scriptsize$\pm$}.0141 \\
     & NDCG@10 & .3834{\scriptsize$\pm$}.0038 & .3540{\scriptsize$\pm$}.0172 & \underline{.4094}{\scriptsize$\pm$}.0099 & .3758{\scriptsize$\pm$}.0088 & .3999{\scriptsize$\pm$}.0063 & .3263{\scriptsize$\pm$}.0127 & .3267{\scriptsize$\pm$}.0158 & .3294{\scriptsize$\pm$}.0059 & \textbf{.4182}{\scriptsize$\pm$}.0108 & .3990{\scriptsize$\pm$}.0058 \\
     & MRR@10 & .3105{\scriptsize$\pm$}.0073 & .2922{\scriptsize$\pm$}.0175 & \underline{.3410}{\scriptsize$\pm$}.0079 & .3053{\scriptsize$\pm$}.0079 & .3248{\scriptsize$\pm$}.0119 & .2736{\scriptsize$\pm$}.0115 & .2756{\scriptsize$\pm$}.0143 & .2835{\scriptsize$\pm$}.0079 & \textbf{.3465}{\scriptsize$\pm$}.0120 & .3207{\scriptsize$\pm$}.0097 \\
    \midrule
    \multirow{3}{*}{Beauty} & HR@10 & .2035{\scriptsize$\pm$}.0271 & .1818{\scriptsize$\pm$}.0225 & \textbf{.2468}{\scriptsize$\pm$}.0260 & \underline{.2425}{\scriptsize$\pm$}.0075 & .2208{\scriptsize$\pm$}.0130 & .1472{\scriptsize$\pm$}.0397 & .1991{\scriptsize$\pm$}.0418 & .1385{\scriptsize$\pm$}.0270 & .2295{\scriptsize$\pm$}.0199 & .2381{\scriptsize$\pm$}.0327 \\
     & NDCG@10 & .1113{\scriptsize$\pm$}.0086 & .1050{\scriptsize$\pm$}.0091 & .1259{\scriptsize$\pm$}.0096 & \textbf{.1395}{\scriptsize$\pm$}.0049 & .1248{\scriptsize$\pm$}.0106 & .0805{\scriptsize$\pm$}.0092 & .1057{\scriptsize$\pm$}.0109 & .0573{\scriptsize$\pm$}.0150 & .1163{\scriptsize$\pm$}.0150 & \underline{.1325}{\scriptsize$\pm$}.0249 \\
     & MRR@10 & .0833{\scriptsize$\pm$}.0036 & .0811{\scriptsize$\pm$}.0077 & .0901{\scriptsize$\pm$}.0053 & \textbf{.1081}{\scriptsize$\pm$}.0075 & .0958{\scriptsize$\pm$}.0114 & .0606{\scriptsize$\pm$}.0016 & .0770{\scriptsize$\pm$}.0036 & .0332{\scriptsize$\pm$}.0109 & .0825{\scriptsize$\pm$}.0128 & \underline{.1009}{\scriptsize$\pm$}.0242 \\
    \midrule
    \multicolumn{2}{l}{\textbf{Avg.~Rank$\downarrow$}} & \multicolumn{1}{c}{4.78} & \multicolumn{1}{c}{6.33} & \multicolumn{1}{c}{4.83} & \multicolumn{1}{c}{5.78} & \multicolumn{1}{c}{5.22} & \multicolumn{1}{c}{7.83} & \multicolumn{1}{c}{6.89} & \multicolumn{1}{c}{7.50} & \multicolumn{1}{c}{4.11} & \multicolumn{1}{c}{\textbf{1.61}} \\
    \bottomrule
  \end{tabular}}
\end{table*}

\subsection{Hierarchical Sparse Expert Allocation}
\label{sec:hierarchical-routing}

RouteRec organizes $G \times K$ experts into $G$ groups of $K$ experts, with one group associated with each cue family (Figure~\ref{fig:framework}). Allocation proceeds in two steps. First, family-specific cue projections score and select expert groups, identifying the behavioral families most relevant to the session. Second, within each selected group, the router selects experts from that family's projected cues and the current backbone state. The resulting path is sparse and behavior-dependent.
Separating these roles keeps group selection tied to explicit behavioral evidence while allowing experts within a group to adapt to the position-specific sequence representation.

\paragraph{Router formulation.}
Let $\mathbf{q}^{(s,g)}_{i,t}$ be the projected cue vector of family $g$ at scope $s$.
For macro and mid routing, cues are pooled over the valid prefix and shared across positions; micro cues remain position-specific.
Let $\mathbf{u}^{(s,g)}_{i,t}=[\widetilde{\mathbf{h}}^{(\ell)}_{i,t};\mathbf{q}^{(s,g)}_{i,t}]$ denote the within-group routing input.
The group and conditional expert weights are
\[
\begin{aligned}
\mathbf{p}^{(s)}_{i,t}
&= \operatorname{TopKSoftmax}_{k_G}\!\left(
  \left\{\phi^{(s)}_g(\mathbf{q}^{(s,g)}_{i,t})\right\}_{g=1}^{G}
\right),
\\
\mathbf{r}^{(s)}_{i,t,g}
&= \operatorname{TopKSoftmax}_{k_E}\!\left(
  \boldsymbol{\psi}^{(s)}_g(\mathbf{u}^{(s,g)}_{i,t})
\right),
\end{aligned}
\]
where $\phi_g$ is a family-specific linear head and $\boldsymbol{\psi}_g$ is a group-specific two-layer network; we use $k_G=3$ selected groups and $k_E=2$ selected experts per group.
$\operatorname{TopKSoftmax}_k$ masks all but the $k$ largest logits and applies softmax over the retained entries.
The joint expert weight is
\[
\pi^{(s)}_{i,t,g,k}
=p^{(s)}_{i,t,g}\,r^{(s)}_{i,t,g,k}.
\]
Thus, macro and mid group selection is session-wise, while their conditional expert weights remain position-wise through the backbone state; both are position-wise at the micro scope.

\paragraph{Sparse selection.}
At both levels, RouteRec masks non-selected entries and renormalizes the active weights. Sessions with different behavioral cues can activate different groups, while sessions sharing a dominant family can still choose different experts through within-family cues and current hidden states. The hierarchy therefore represents behavioral variation at both coarse family and fine expert levels.

\subsection{Training Objective}
\label{sec:training-objective}

RouteRec is trained end-to-end with the next-item prediction loss and two routing regularizers.
The regularizers stabilize the cue-to-route mapping and routing logits while preserving behavior-dependent expert concentration.

\paragraph{Prediction loss.}
The main objective is standard next-item cross-entropy.
Let $\mathbf{z} \in \mathbb{R}^{|\mathcal{I}|}$ be the output logit vector over all candidate items $\mathcal{I}$.
Let $z_y$ denote the logit for the target item $y$:
\begin{equation}
\mathcal{L}_{\mathrm{CE}} = -\log \frac{\exp(z_y)}{\sum_{j \in \mathcal{I}}\exp(z_j)}.
\end{equation}

\paragraph{Route consistency.}
The route consistency term encourages the cue-to-route mapping to vary smoothly across behaviorally similar sessions.
Let $\mathcal{T}=\{\text{macro},\text{mid},\text{micro}\}$ denote the set of routing scopes.
Let $\mathbf{c}_i$ be the $\ell_2$-normalized, position-averaged cue representation of session $i$, and let $k_B=\min(4,B-1)$ for a batch of size $B$.
We form the directed neighbor set $\mathcal{N}_{k_B}(i)$ from the $k_B$ most cosine-similar non-self sessions in the batch and reuse it across all routing scopes.
Let $\bar{\boldsymbol{\pi}}^{(s)}_i$ denote the valid-position mean routing distribution of session $i$ at scope $s$.
The loss penalizes Jensen--Shannon divergence between each session and its selected neighbors:
\[
\mathcal{L}_{\mathrm{cons}}
= \frac{1}{|\mathcal{T}|Bk_B}
\sum_{s \in \mathcal{T}}\sum_{i=1}^{B}
\sum_{j \in \mathcal{N}_{k_B}(i)}
\mathrm{JSD}\!\left(
\bar{\boldsymbol{\pi}}^{(s)}_i,
\bar{\boldsymbol{\pi}}^{(s)}_j
\right).
\]

\paragraph{Routing logit stabilization.}
The sparse routing of RouteRec uses only a small expert subset.
Large pre-softmax logits can make the resulting weights overly concentrated.
We thus apply z-loss~\cite{zoph2022stmoe} to penalize large logit magnitudes:
\[
\mathcal{L}_{z}
= \frac{1}{|\mathcal{T}|} \sum_{s \in \mathcal{T}} \frac{1}{|\mathcal{V}_s|}
\sum_{(i,t)\in\mathcal{V}_s}
\!\left(\log \sum_{g,k}\exp a^{(s)}_{i,t,g,k}\right)^{\!2},
\]
where $a^{(s)}_{i,t,g,k}$ is the final joint pre-softmax logit produced by the factored group and within-group router for session $i$, position $t$, group $g$, and expert $k$ at scope $s$; masked pairs are excluded.
$\mathcal{V}_s$ is the set of valid (session, position) pairs used for computing the loss term.

\paragraph{Overall objective.}
The total objective is
\[
\mathcal{L} = \mathcal{L}_{\mathrm{CE}} + \lambda_{\mathrm{cons}}\,\mathcal{L}_{\mathrm{cons}} + \lambda_{z}\,\mathcal{L}_{z},
\]
where $\lambda_{\mathrm{cons}}$ and $\lambda_z$ are regularization weights.
Route consistency and z-loss together stabilize behavior-guided routing while preserving behavior-dependent expert concentration.

\section{Experiments}
\label{sec:experiments}

We evaluate RouteRec through three research questions.
\textbf{RQ1}: Does behavior-guided expert allocation improve next-item prediction across diverse datasets?
\textbf{RQ2}: Do behavioral cues control the learned computation path rather than merely adding capacity?
\textbf{RQ3}: Which routing signals, scopes, expert organizations, and objectives support this effect?
We answer them with overall comparisons, routing analyses, and targeted design ablations.

\subsection{Experimental Setup}

\paragraph{Datasets.}
We predict the last item of each held-out session on six public datasets with different domains and session patterns: KuaiRec~\cite{gao2022kuairec}, LastFM~\cite{celma2010music}, Retail Rocket~\cite{retailrocket2015dataset}, ML-1M~\cite{harper2015movielens}, Foursquare~\cite{yang2015foursquare}, and Beauty~\cite{mcauley2015amazon}. Each uses source-appropriate session construction and a common chronological protocol. Appendix~\ref{app:data-details} reports processed statistics; exact rules and frozen manifests are released with the code.

\paragraph{Metrics and reporting.}
Under a shared top-10 protocol, we report hit rate (HR@10), normalized discounted cumulative gain (NDCG@10), and mean reciprocal rank (MRR@10). Targets absent from training are excluded for every method, giving all models the same closed candidate catalog~\cite{latifi2021sessionaware}. We select hyperparameters and checkpoints by the mean of validation HR@10, NDCG@10, and MRR@10, then evaluate the test set once. Results report mean and sample standard deviation over seeds 42--44.

\paragraph{Baselines.}
Nine baselines cover three complementary groups. \textit{Conventional baselines} (GRU4Rec~\cite{hidasi2016gru4rec}, SASRec~\cite{kang2018sasrec}, and TiSASRec~\cite{li2020tisasrec}) use recurrent or self-attentive sequence modeling. \textit{Representation-enhanced baselines} (DuoRec~\cite{qiu2022duorec}, FEARec~\cite{du2023fearec}, and BSARec~\cite{shin2024bsarec}) add contrastive or frequency-aware objectives. \textit{Side-information and MoE baselines} (FDSA~\cite{zhang2019fdsa}, DIF-SR~\cite{xie2022difsr}, and FAME~\cite{liu2025fame}) incorporate item attributes or expert decomposition. All methods use the same processed data, optimizer family, and selection rule.

\subsection{Overall Performance}

RouteRec obtains the best overall Avg.~Rank of 1.61 across the 18 dataset--metric cells, compared with 4.11 for the next-best baseline, FDSA (Table~\ref{tab:main-results}).
It ranks first in 12 cells and second in three, and achieves the best HR@10, NDCG@10, and MRR@10 on KuaiRec, LastFM, and Retail Rocket.
The clearest margins occur on LastFM: RouteRec improves NDCG@10 from .2703 to .2866 over the next-best SASRec, a relative gain of 6.0\%. LastFM's long listening histories expose tempo, repetition, and item-group focus across multiple time scales.
On KuaiRec and Retail Rocket, the improvements are smaller but consistent across all three metrics. KuaiRec sessions mix exploration with concentrated re-watching, which yields several routing signals, although platform exposure may also shape them. On Retail Rocket, category, repetition, and novelty cues still separate browsing from repeat-purchase behavior despite shorter sessions.

We interpret the remaining dataset-level patterns descriptively; the comparison does not isolate causal properties. On ML-1M, RouteRec ranks first on NDCG@10 and MRR@10 and narrowly second on HR@10; its advantage is clearer for ordering retrieved targets than for expanding top-10 coverage. Rating-derived sessions express session-level behavioral regimes less directly than natural interaction sessions.
On Foursquare, RouteRec achieves the best HR@10 but ranks fourth on NDCG@10 and MRR@10; the small, short-session corpus provides limited support for learning fine-grained routing patterns.
On Beauty, RouteRec ranks second on NDCG@10 and MRR@10 and third on HR@10, with relatively large variation across seeds; Beauty's sparse, short histories provide less stable prefix-level behavioral cues.
Aggregate performance alone also does not show whether behavioral cues affect expert allocation, so the next subsection examines the routing mechanism directly.

\subsection{Evidence for Behavior-Guided Routing}
\label{sec:routing-analysis}

We next examine whether the gains are accompanied by behavior-dependent expert allocation. Routing profiles test whether observed behavior changes expert usage, while cue perturbation tests whether the router uses the meaning of its cue input. Capacity, backbone, and training-time routing-source controls then separate this effect from shared computation, hidden-state routing, and misaligned cues. The checks therefore move from observed allocation patterns to increasingly direct controls of the routing mechanism.

\begin{figure}[t]
  \centering
  \NewFig[width=0.78\columnwidth]{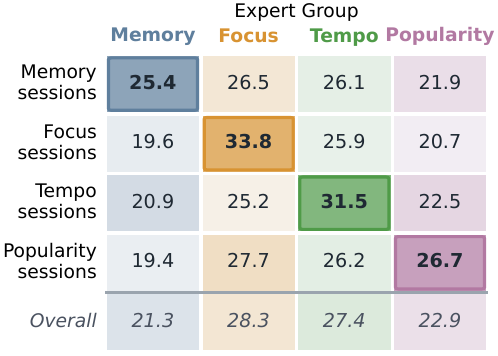}
  \caption{Routing profiles on KuaiRec, measured as scope-averaged route share (\%). Sessions are partitioned by dominant behavioral family; the highlighted diagonal indicates each family's matched expert group.}
  \Description{Routing profile alignment matrix on KuaiRec, with Overall row.}
  \label{fig:routing-profiles}
\end{figure}

\paragraph{Routing Profiles.}
We first ask whether allocation varies with a session's dominant behavioral family, defined as the family whose cue values deviate most from that family's dataset-wide distribution. On KuaiRec, we partition test sessions by this family and average each group's selection probability over valid positions and scopes (Figure~\ref{fig:routing-profiles}); each row sums to 100\%. Relative to the Overall row, every family increases the share of its matched group by 3.8--5.5 percentage points. Thus, despite nonuniform overall expert usage, routing shifts consistently toward the group associated with the dominant cues. This pattern is consistent with behavior-aligned specialization beyond global usage preferences.

\paragraph{Evaluation-Time Cue Perturbation.}
Cue perturbation tests whether the trained router relies on cue semantics at inference time. We corrupt a trained, frozen model while leaving backbone and expert weights unchanged. \textit{Zero cues} asks whether cue information is needed at all, whereas \textit{Shuffle cues} preserves realistic cue vectors but assigns them to the wrong sessions. \textit{Global permute} scrambles values within each family. \textit{Role swap} exchanges entire vectors between families, preserving their magnitudes while assigning them the wrong behavioral role.

\begin{figure}[t]
  \centering
  \NewFig[width=\columnwidth]{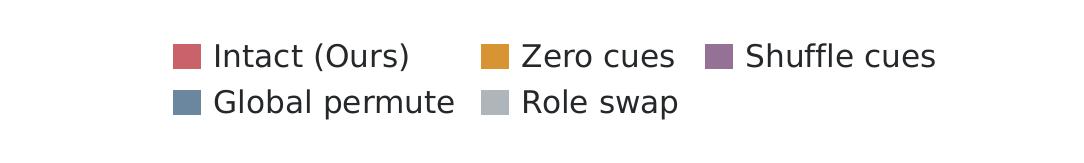}
  \vspace{0.25em}
  \subcaptionbox{KuaiRec}{%
    \NewFig[width=0.485\columnwidth]{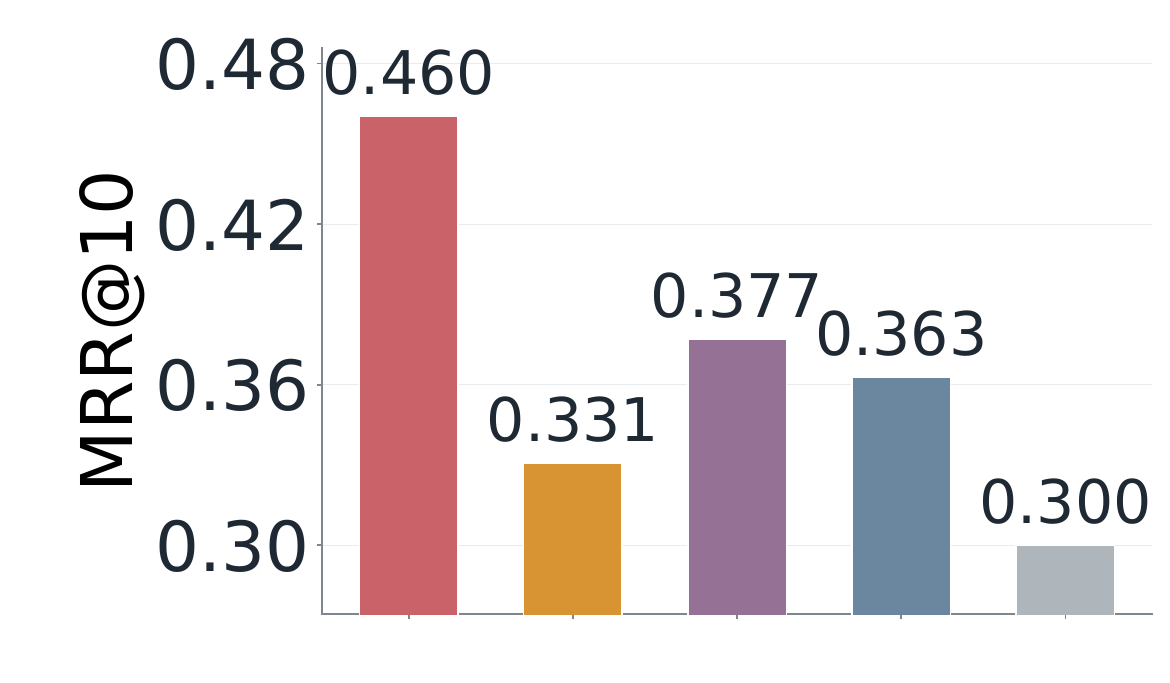}%
  }\hfill
  \subcaptionbox{LastFM}{%
    \NewFig[width=0.485\columnwidth]{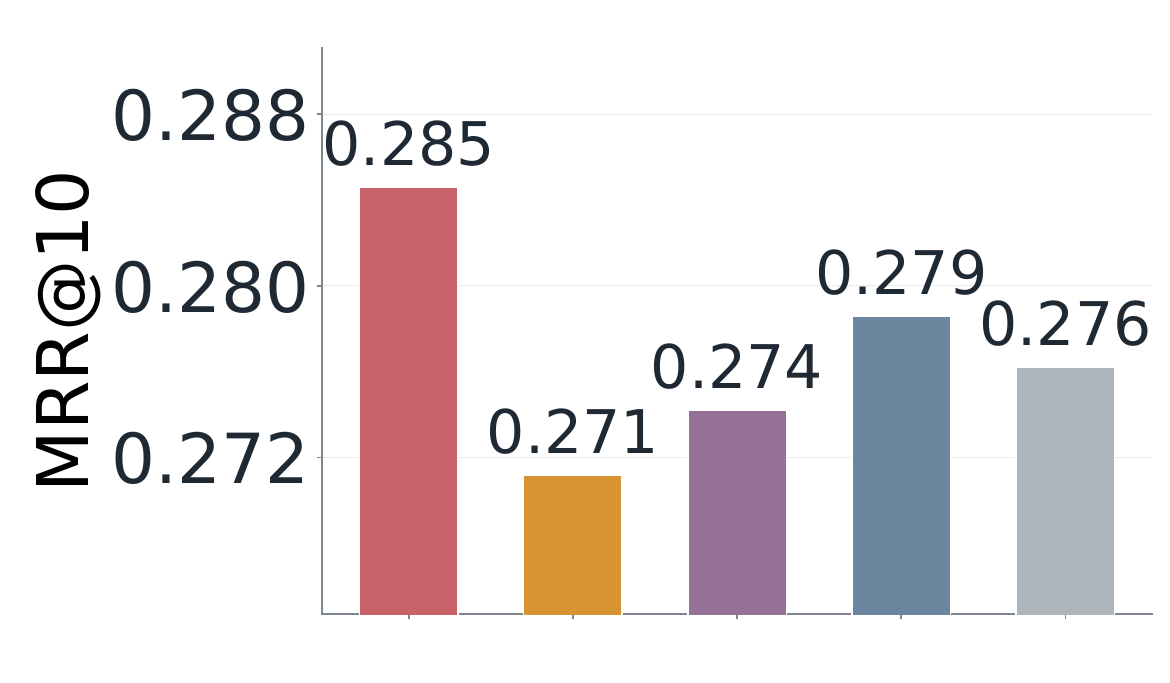}%
  }
  \caption{Evaluation-time cue perturbation results (MRR@10). A trained, frozen RouteRec is evaluated under four cue-corruption strategies to test whether the router relies on cue semantics at inference.}
  \Description{MRR@10 under cue corruptions on KuaiRec and LastFM.}
  \label{fig:cue-perturbation-main}
\end{figure}

All corruptions reduce MRR@10 on both datasets (Figure~\ref{fig:cue-perturbation-main}), showing that the trained model uses its behavioral cue input. On KuaiRec, role swap is more damaging than zeroing cues (approximately 35\% vs.\ 28\% relative MRR@10 reduction), so a structurally plausible but behaviorally misaligned cue can be worse than no cue. LastFM shows smaller but consistent reductions, although the corruption ranking differs. Together, the results support cue-dependent routing on both datasets and stronger cue--family sensitivity on KuaiRec.

\begin{figure}[!t]
  \centering
  \NewFig[width=\columnwidth]{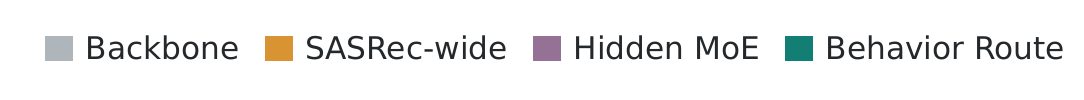}
  \subcaptionbox{SASRec attribution}{%
    \NewFig[width=0.485\columnwidth]{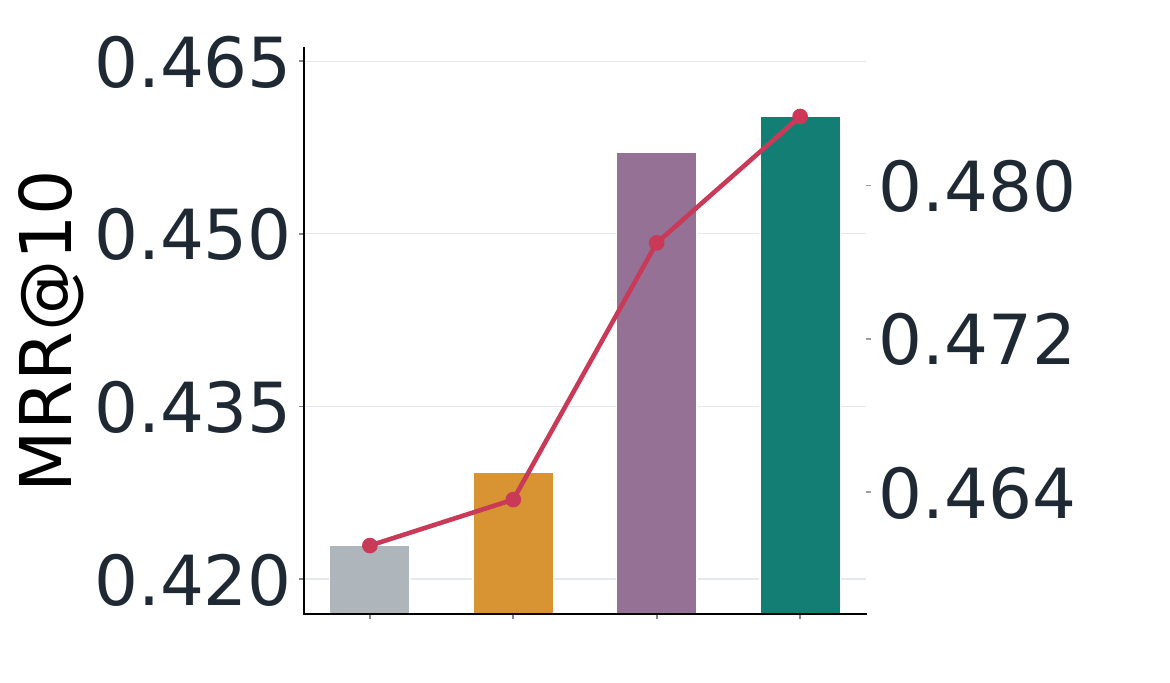}%
  }\hfill
  \subcaptionbox{DuoRec portability}{%
    \NewFig[width=0.485\columnwidth]{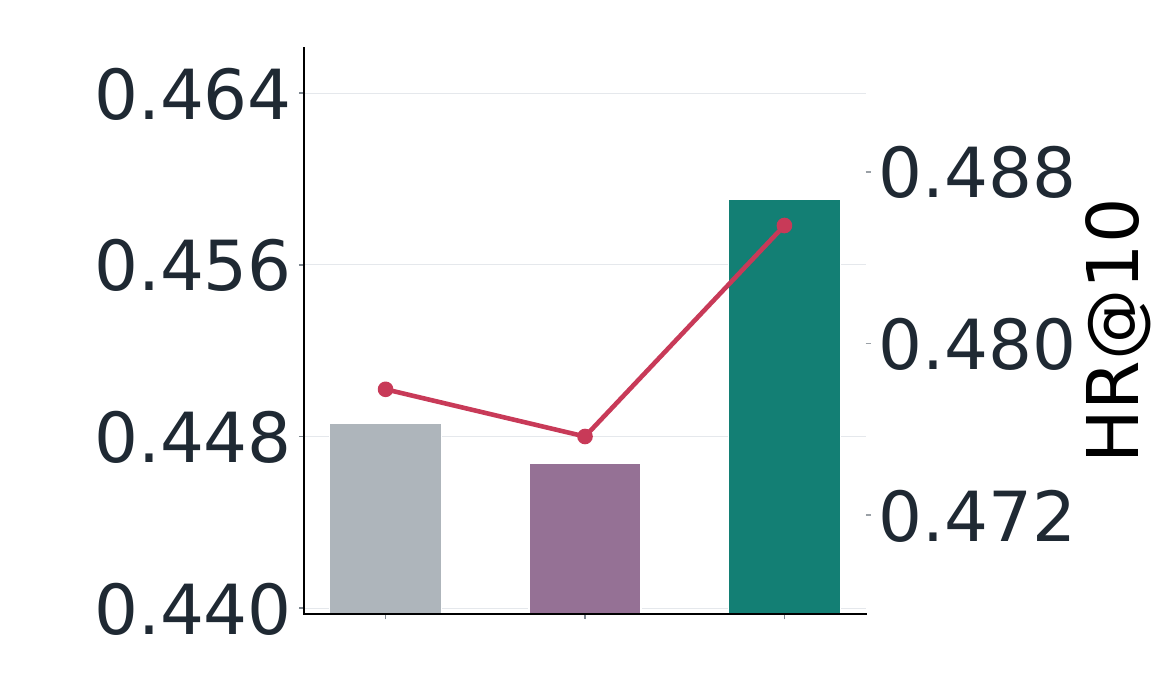}%
  }
  \caption{Routing attribution and backbone portability on KuaiRec. Bars show MRR@10 and lines show HR@10: (a) SASRec-based attribution controls; (b) matched DuoRec variants. \textit{Backbone} denotes SASRec in (a) and DuoRec in (b), and SASRec-wide appears only in (a). The y-axes are truncated to make differences visible.}
  \Description{Two KuaiRec panels with MRR at 10 as bars and HR at 10 as lines. The left panel compares SASRec, SASRec-wide, Hidden MoE, and Behavior Route. The right panel compares DuoRec, DuoRec with Hidden MoE, and DuoRec with Behavior Route.}
  \label{fig:backbone-routing-control}
\end{figure}

\paragraph{Routing Attribution and Backbone Portability.}
Figure~\ref{fig:backbone-routing-control} uses two comparisons on KuaiRec. Panel (a) compares SASRec, parameter-matched SASRec-wide (within 0.4\%; Appendix~\ref{app:efficiency}), Hidden MoE with the same expert structure, and RouteRec. This separates shared capacity from the source used for routing. Panel (b) applies the routed stages to DuoRec while retaining its contrastive objectives, with vanilla and Hidden-MoE variants as backbone and routing-source controls. Behavior-guided variants are labeled \textit{Behavior Route}.

Widening SASRec yields only a small gain, whereas both routed variants improve MRR@10 and HR@10; RouteRec performs best, though Hidden MoE remains a strong control (Figure~\ref{fig:backbone-routing-control}a). Under DuoRec, behavior-guided routing outperforms both vanilla DuoRec and DuoRec with Hidden MoE on both metrics, while Hidden MoE does not improve the backbone (Figure~\ref{fig:backbone-routing-control}b). The benefit therefore extends beyond SASRec and is not explained by shared capacity or an MoE block alone, although its magnitude depends on the backbone.

\paragraph{Training-Time Routing Source.}
The preceding analyses examine a trained router at evaluation time and compare routing mechanisms across backbones. We next test which routing signals are useful during training and whether cue--session alignment matters. \textit{Cue only} and \textit{Hidden only} each retain a single routing source, while \textit{Train shuffle} and \textit{Train zero} respectively misalign and remove cues.

\begin{figure}[!b]
  \centering
  \NewFig[width=\columnwidth]{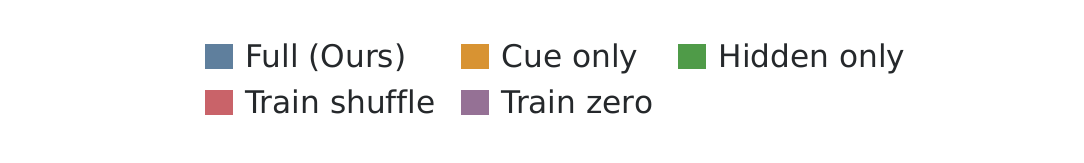}\\[2pt]
  \subcaptionbox{KuaiRec}{%
    \NewFig[width=0.485\columnwidth]{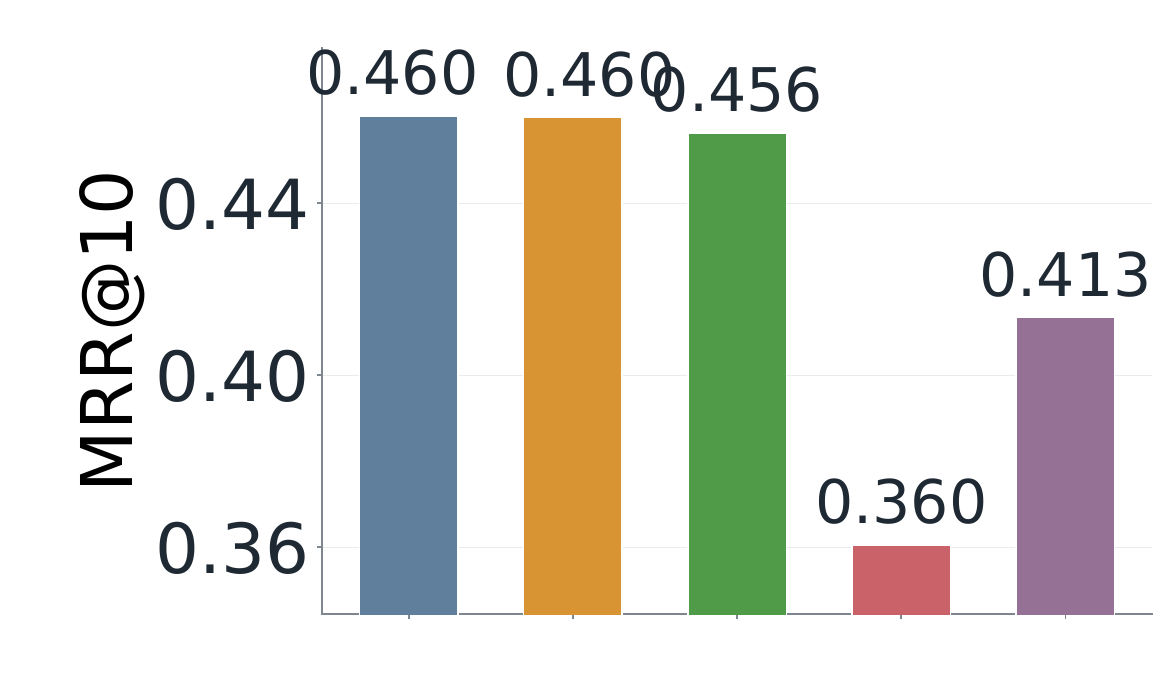}%
  }\hfill
  \subcaptionbox{LastFM}{%
    \NewFig[width=0.485\columnwidth]{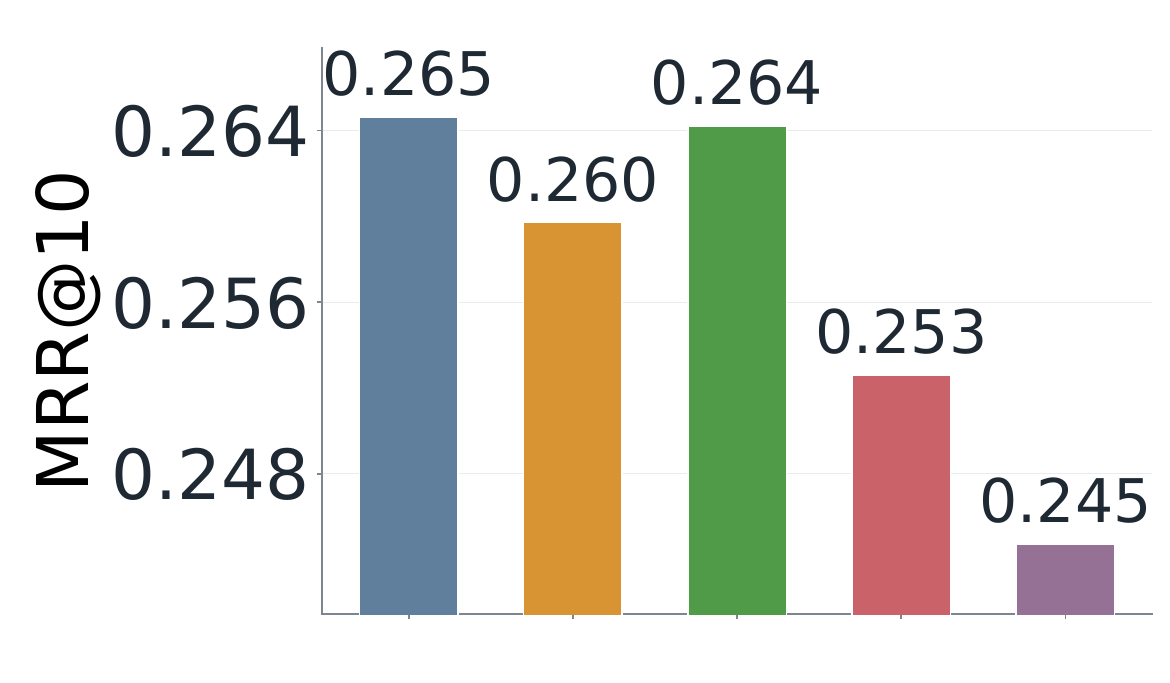}%
  }
  \caption{Train-time routing-source ablation results (MRR@10). Clean cue-only and hidden-only controls are compared with variants that misalign or remove cues during training, testing which signals support routing and whether cue--session alignment matters.}
  \Description{MRR at 10 for full, cue-only, hidden-only, train-shuffle, and train-zero routing variants on KuaiRec and LastFM.}
  \label{fig:train-perturb-main}
\end{figure}

Both single-source variants remain close to the full model (Figure~\ref{fig:train-perturb-main}), although their relative strengths differ across datasets. Cue corruption is more damaging: \textit{Train shuffle} sharply degrades performance on KuaiRec, while \textit{Train zero} falls to approximately SASRec level on both datasets. Thus, coherent cue--session alignment matters during training, with hidden states providing a complementary signal. Appendix~\ref{app:cue-availability} evaluates RouteRec when whole cue families are unavailable.

\subsection{Core Architectural Choices}
\label{sec:ablation}

Having established that routing responds to behavioral evidence, we next examine the architectural choices that implement it: the number and contribution of routing scopes, the organization of experts, and the auxiliary routing objectives. These ablations test whether the effect requires the complete cue-to-route design or can be reproduced by a simpler routing structure.

\begin{figure}[t]
  \centering
  \begin{subfigure}[t]{0.49\columnwidth}
    \centering
    \NewFig[width=\linewidth]{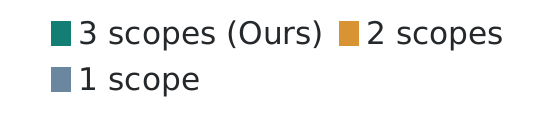}\\[-2pt]
    \NewFig[width=\linewidth]{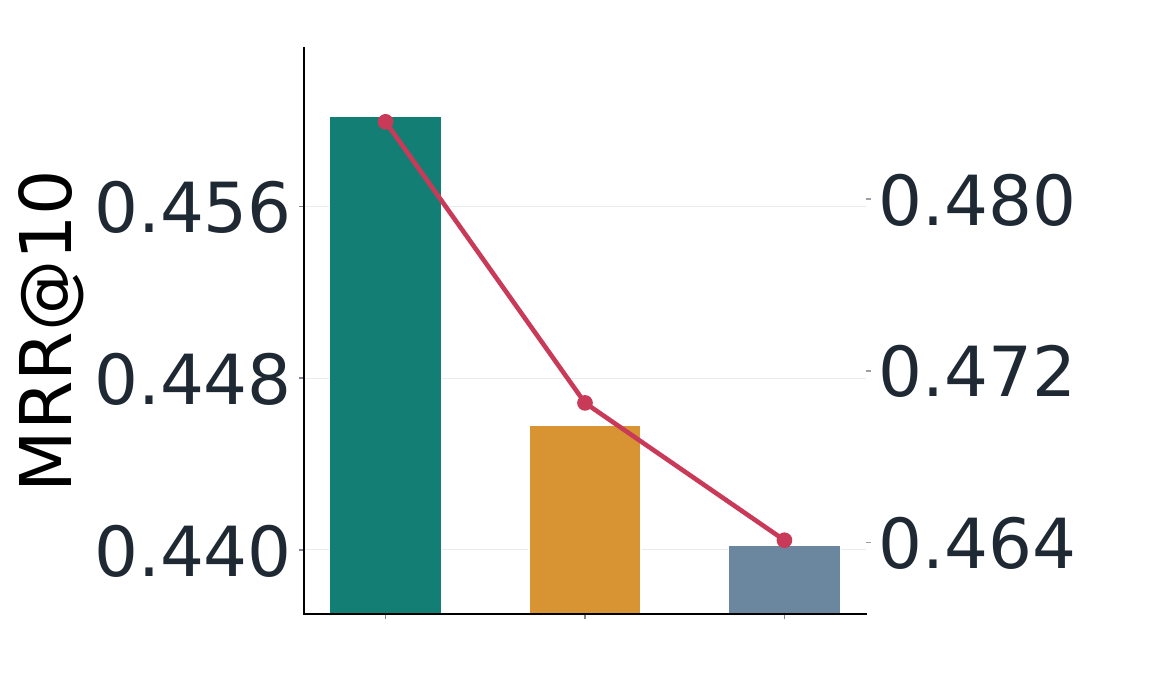}
    \caption{Temporal scopes}
  \end{subfigure}\hfill
  \begin{subfigure}[t]{0.49\columnwidth}
    \centering
    \NewFig[width=\linewidth]{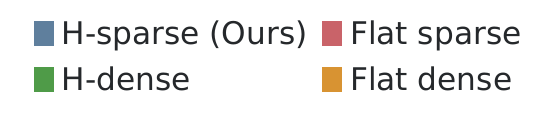}\\[-2pt]
    \NewFig[width=\linewidth]{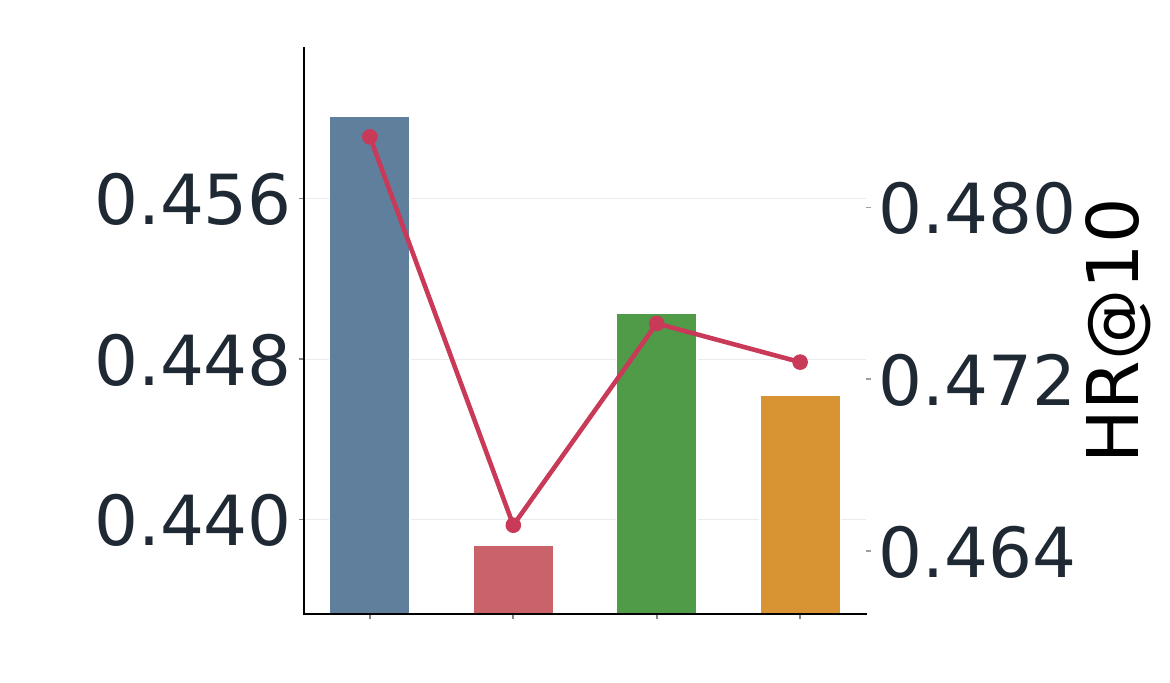}
    \caption{Routing organization}
  \end{subfigure}
  \caption{Design ablations on KuaiRec (MRR@10 bars, HR@10 lines). \textit{Left}: scope-count variants test whether more routing scopes consistently help. \textit{Right}: routing-organization variants compare hierarchical sparse against flatter or denser alternatives.}
  \Description{Ablations for temporal routing scopes and routing organization.}
  \label{fig:stage-structure-panels-main}
\end{figure}

\paragraph{Routing Scopes and Organization.}
Figure~\ref{fig:stage-structure-panels-main} tests two structural choices on KuaiRec.
We compare the full three-scope model with two- and one-scope variants, and the hierarchical sparse router with hierarchical dense, flat sparse, and flat dense alternatives.

Performance declines progressively as the number of scopes is reduced (Figure~\ref{fig:stage-structure-panels-main}, left), supporting routing over multiple behavioral time scales.
Hierarchical sparse routing performs best, and both hierarchical variants outperform their flat counterparts (Figure~\ref{fig:stage-structure-panels-main}, right), indicating that cue-family organization contributes beyond sparse expert selection alone.

\paragraph{Scope Contributions.}
The scope-count ablation tests how many routing scopes to use in total.
As a complement, we remove each scope individually while keeping the other two active, isolating each scope's contribution to the full macro--mid--micro stack.

\begin{figure}[t]
  \centering
  \NewFig[width=\columnwidth]{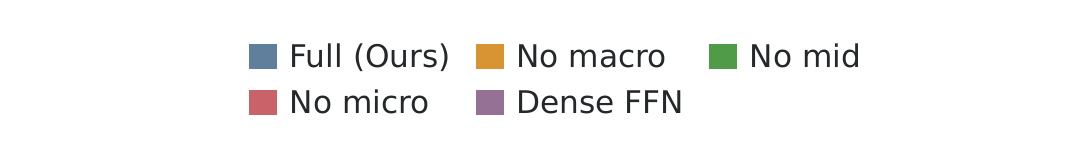}\\[2pt]
  \subcaptionbox{KuaiRec}{%
    \NewFig[width=0.485\columnwidth]{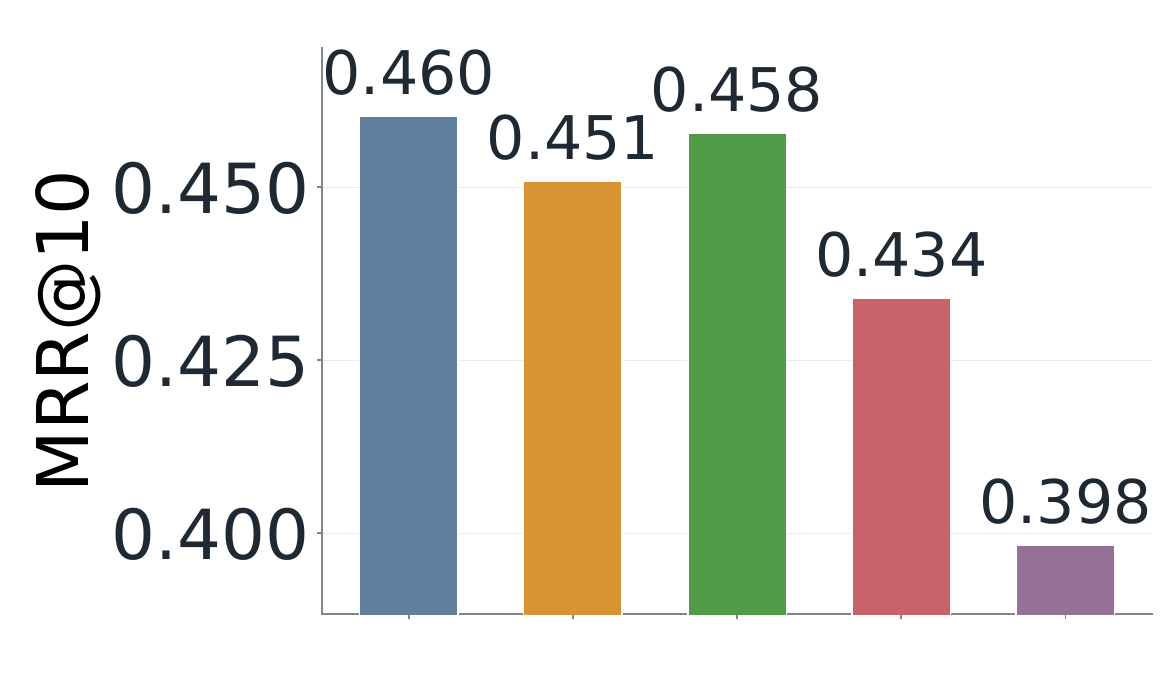}%
  }\hfill
  \subcaptionbox{LastFM}{%
    \NewFig[width=0.485\columnwidth]{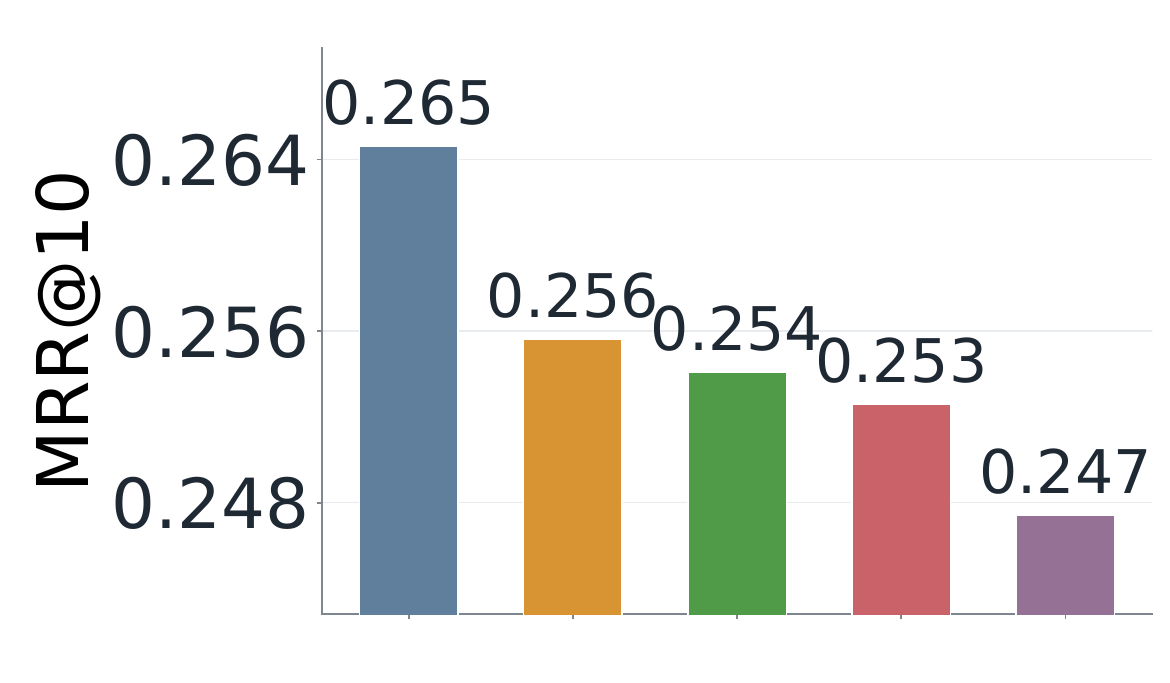}%
  }
  \caption{Scope contribution ablation (MRR@10). Each routing scope is removed from the full macro--mid--micro stack in turn, with the other two remaining active, to isolate each scope's individual contribution; a dense feed-forward baseline is included for reference.}
  \Description{Scope contribution ablation on KuaiRec and LastFM.}
  \label{fig:core-stage-main}
\end{figure}

Removing any scope lowers MRR@10 on both datasets (Figure~\ref{fig:core-stage-main}).
On KuaiRec, removing the micro scope causes the largest decline by a clear margin; on LastFM, the effects are closer, with the no-mid and no-micro variants yielding nearly identical MRR@10.
The dense feed-forward baseline performs worst on both datasets, indicating that routed computation contributes beyond simply widening the shared feed-forward block.
The dataset difference also suggests that the most useful temporal resolution depends on how behavior unfolds, even though all three scopes contribute to the complete model.

\paragraph{Auxiliary Losses.}
RouteRec adds route consistency and z-loss to the cross-entropy objective (Section~\ref{sec:training-objective}).
The former aligns routing distributions for sessions with similar cues; the latter controls routing-logit scale.
We ablate each and both, and compare standard load balancing.

\begin{figure}[t]
  \centering
  \NewFig[width=\columnwidth]{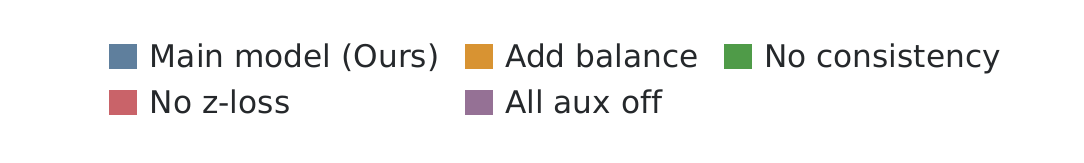}\\[2pt]
  \subcaptionbox{KuaiRec}{%
    \NewFig[width=0.485\columnwidth]{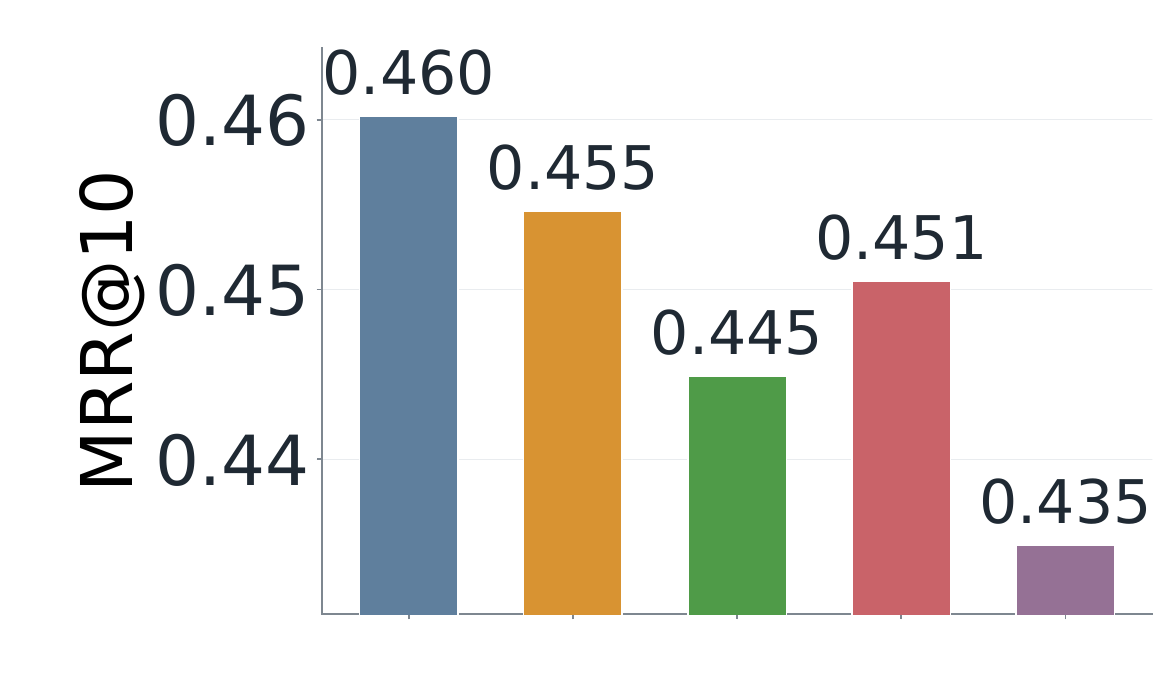}%
  }\hfill
  \subcaptionbox{LastFM}{%
    \NewFig[width=0.485\columnwidth]{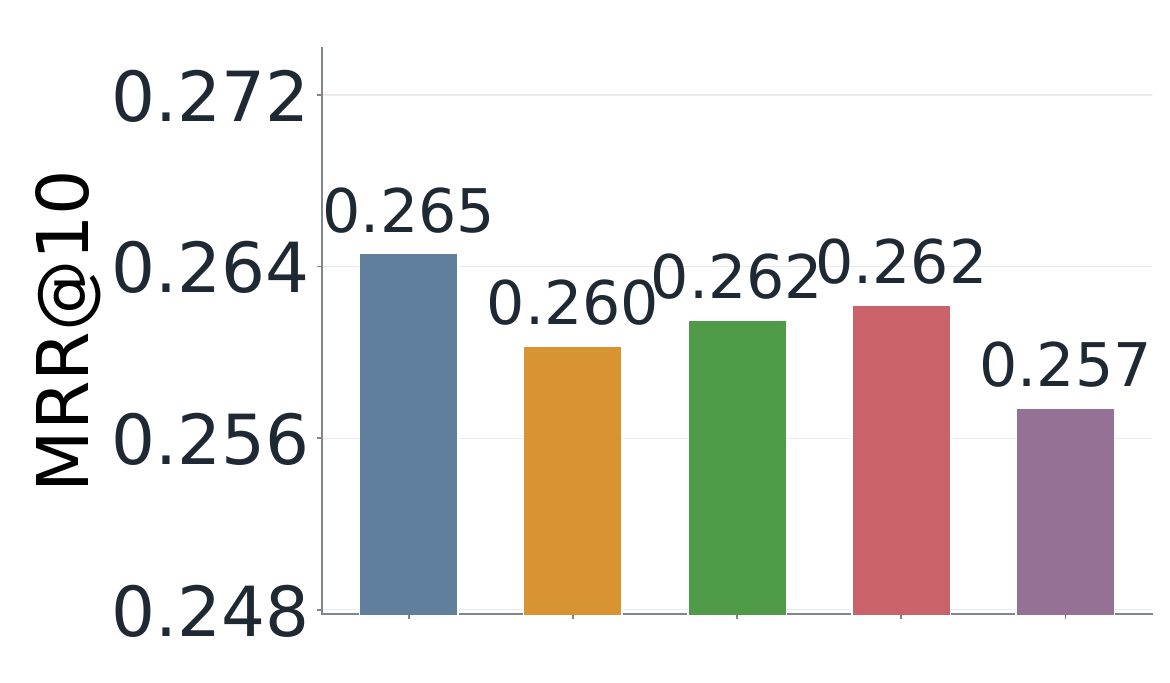}%
  }
  \caption{Auxiliary-loss ablation results (MRR@10). We compare the full objective with variants that remove route consistency, z-loss, or both, and with a variant that adds load-balancing regularization.}
  \Description{Auxiliary-loss ablation on KuaiRec and LastFM.}
  \label{fig:aux-loss-main}
\end{figure}

Removing both regularizers causes the largest drop on both datasets (Figure~\ref{fig:aux-loss-main}).
Route consistency matters more than z-loss on KuaiRec and about as much on LastFM.
Load balancing also degrades performance on both datasets, so we omit it: uniform aggregate usage conflicts with behavior-dependent allocation.

\section{Conclusion}
\label{sec:conclusion}

We present RouteRec, a hierarchical sparse MoE recommender that routes sessionized histories using prefix-derived Tempo, Focus, Memory, and Popularity cues across macro, mid, and micro scopes.
Across six datasets and nine baselines, RouteRec achieves the best overall average rank and ranks first in 12 of the 18 dataset--metric cells.
Routing profiles and cue perturbations show that expert allocation shifts with observed behavior and depends on cue--session alignment.
The SASRec controls indicate that the gains are not explained by shared capacity alone, while the KuaiRec DuoRec comparison shows that behavior-guided routing can transfer beyond the SASRec backbone.

The dataset-level variation is consistent with, but does not establish, the hypothesis that routing is most useful when session prefixes reveal distinct behavior regimes with sufficient training support.
RouteRec should therefore be viewed as a method for heterogeneous sessionized histories rather than a uniform improvement across all sequential datasets.
The current dense expert execution is costlier than SASRec-wide (Appendix~\ref{app:efficiency}), and some cues require timestamps or item-group metadata; conditional expert execution and more robust cue construction are natural directions for future work.

\begin{acks}
This work was partly supported by the Institute of Information \& Communications Technology Planning \& Evaluation (IITP) grant funded by the Ministry of Science and ICT (MSIT) under Grant No. RS-2026-25526850, by the National Research Foundation of Korea (NRF) grant funded by the Korea government (MSIT) (RS-2024-00341425, RS-2024-00406985), by the “Advanced GPU Utilization Support Program” funded by the Government of the Republic of Korea (Ministry of Science and ICT), and by the New Faculty Startup Fund from Seoul National University.
\end{acks}

\appendix

\section{Cue Availability under Limited Metadata}
\label{app:cue-availability}

To test robustness to unavailable metadata, we retrain RouteRec under three reduced-cue conditions. \textit{No category cues} removes label-dependent Focus cues, \textit{No time cues} removes timestamp-dependent Tempo cues, and \textit{Sequence-only cues} removes both families. We compare each variant with the full model and the best non-RouteRec baseline.

\begin{figure}[t]
  \centering
  \NewFig[width=\columnwidth]{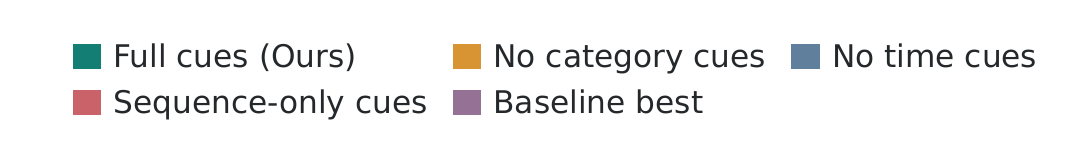}
  \vspace{0.25em}
  \subcaptionbox{KuaiRec}{%
    \NewFig[width=0.485\columnwidth]{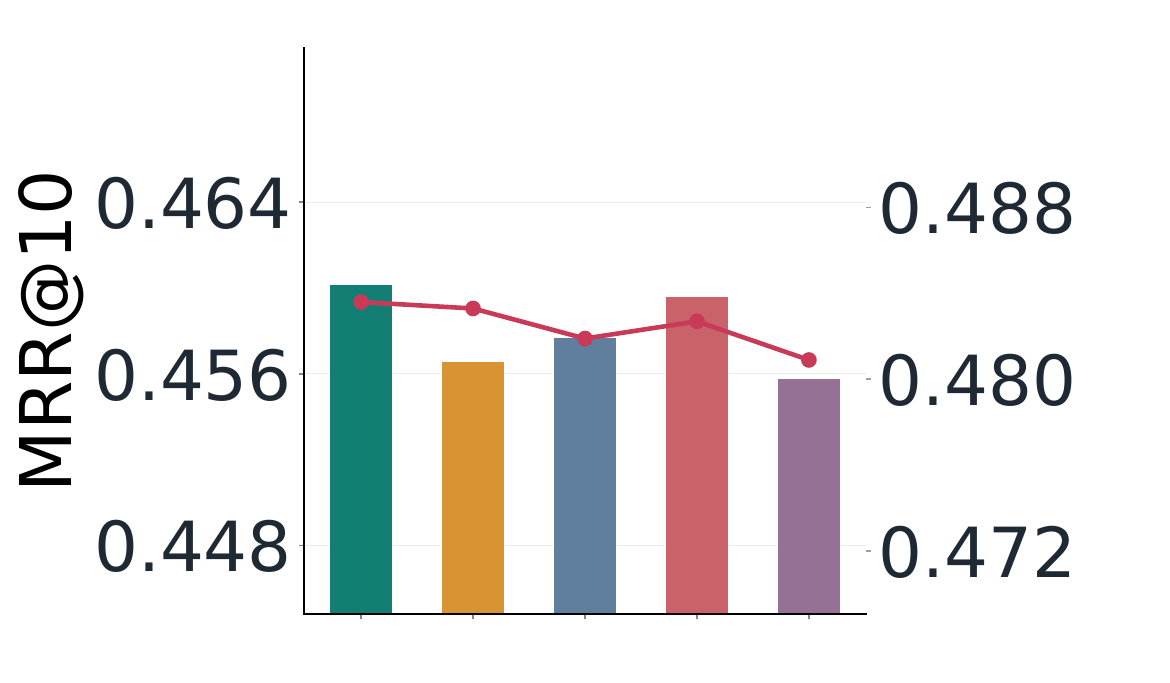}%
  }\hfill
  \subcaptionbox{LastFM}{%
    \NewFig[width=0.485\columnwidth]{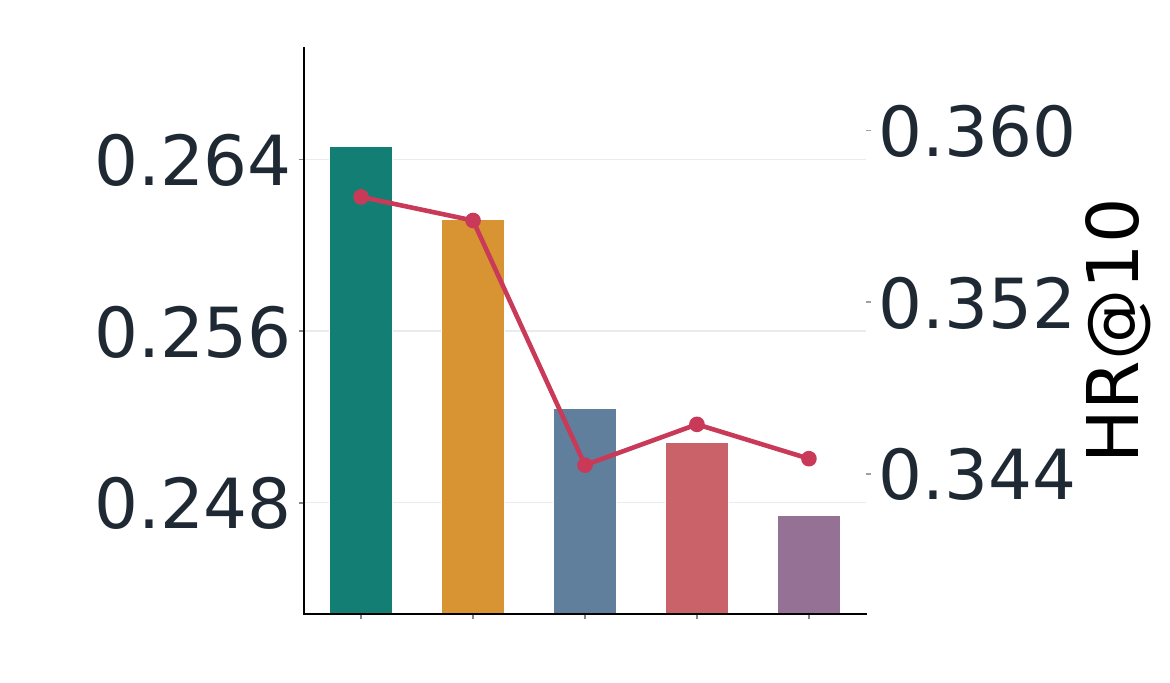}%
  }
  \caption{Cue-family ablation results (MRR@10 bars, HR@10 line). RouteRec is retrained under three reduced-cue conditions to test whether behavior-guided routing remains effective when metadata sources are partially unavailable.}
  \Description{Dual-axis bar and line charts comparing RouteRec under full and reduced cue conditions.}
  \label{fig:cue-family-ablation}
\end{figure}

Cue availability has a dataset-dependent effect (Figure~\ref{fig:cue-family-ablation}).
On KuaiRec, all reduced-cue variants remain close to the full model, including the sequence-only variant, showing that item-identity and frequency cues still provide a useful routing signal.
LastFM is more sensitive to temporal information: removing Tempo cues, either alone or together with Focus cues, causes a larger decline, whereas removing Focus cues alone has a smaller effect.
All reduced-cue variants nevertheless remain competitive with the strongest non-RouteRec baseline, indicating that RouteRec degrades gracefully when timestamps or item-group labels are unavailable.

\section{Implementation Details}
\label{app:method-details}

\subsection{RouteRec Instantiation}

Table~\ref{tab:appendix-main-setting} records the fixed routing layout and activation settings needed to reproduce RouteRec. Router layout and activation are fixed across datasets; capacity and optimization values are selected per dataset from Appendix~\ref{app:data-details}.

\begin{table}[t]
  \centering
  \caption{RouteRec hyperparameter configuration used in the main experiments. The router layout and activation pattern are fixed across datasets; capacity and optimization values are tuned per dataset.}
  \label{tab:appendix-main-setting}
  \footnotesize
  \setlength{\tabcolsep}{3pt}
  \renewcommand{\arraystretch}{1.10}
  \begin{tabular}{@{}p{0.31\columnwidth}p{0.65\columnwidth}@{}}
    \toprule
    \textbf{Setting} & \textbf{Value} \\
    \midrule
    Backbone layout & self-attn $\to$ macro block $\to$ mid block $\to$ self-attn $\to$ micro block \\
    Routing granularity & Macro/mid group gate: session-wise; conditional expert gate: position-wise; micro: both position-wise \\
    Expert groups & $G=4$ (Tempo, Focus, Memory, Popularity) \\
    Experts per group & $K$ selected per dataset from $\{3,4,5,6\}$ \\
    Expert activation & Top-$k_G{=}3$ groups; top-$k_E{=}2$ experts per selected group (6 active) \\
    Cue dimensionality & 16 scalars per scope; 4 families $\times$ 4 summaries each \\
    History windows & $H=5$ sessions (macro); $w=5$ items (micro) \\
    Consistency pairs & Four most similar non-self batch sessions in mean-cue cosine space; directed neighbors shared across scopes \\
    Missing signals & Zero-fill Focus cues when grouping labels are absent \\
    Auxiliary losses & Route consistency + z-loss; no load balancing \\
    \bottomrule
  \end{tabular}
\end{table}

Macro and mid cues are computed once per session and shared across routed positions, whereas micro cues are re-evaluated at each position from the last $w$ items. Group- and expert-level top-$k$ weights are renormalized over the active support. All cues use only observed history and training-split statistics; no held-out target information enters the router.

\subsection{Cue Bank by Scope and Family}
\label{app:cue-bank}

Table~\ref{tab:appendix-feature-families} lists the complete cue bank introduced in Section~\ref{sec:feature-construction}.

\begin{table*}[t]
  \centering
  \caption{Full cue bank used by RouteRec, grouped by family and routing scope. Macro uses the five-session history window; mid uses the observed active session; micro uses the recent local window.}
  \label{tab:appendix-feature-families}
  \scriptsize
  \setlength{\tabcolsep}{3pt}
  \renewcommand{\arraystretch}{1.15}
  \begin{tabular}{@{}p{0.105\textwidth}p{0.285\textwidth}p{0.285\textwidth}p{0.285\textwidth}@{}}
    \toprule
    \textbf{Family} &
    \textbf{Macro} \textit{(cross-session, H=5)} &
    \textbf{Mid} \textit{(active session)} &
    \textbf{Micro} \textit{(recent $w$ items)} \\
    \midrule
    \textbf{Tempo}
      & \cuecell{valid-history ratio}{inter-session gap}{mean pace}{pace trend}
      & \cuecell{valid active prefix}{mean interval}{interval std.}{session age}
      & \cuecell{valid local suffix}{last gap}{mean local gap}{gap shift vs. mid} \\[2pt]
    \textbf{Focus}
      & \cuecell{theme entropy}{top-theme mass}{theme repeat}{theme shift}
      & \cuecell{category entropy}{top-category mass}{switch rate}{category uniq.}
      & \cuecell{current switch}{last-category mismatch}{suffix cat. entropy}{suffix cat. uniq.} \\[2pt]
    \textbf{Memory}
      & \cuecell{repeat intensity}{adj.-category overlap}{adj.-item overlap}{repeat trend}
      & \cuecell{item uniqueness}{repeat rate}{novel-item rate}{longest item run}
      & \cuecell{last reconsumed}{suffix reconsume rate}{suffix item uniq.}{suffix longest run} \\[2pt]
    \textbf{Popularity}
      & \cuecell{mean popularity}{pop. variability}{pop. entropy}{pop. trend}
      & \cuecell{mean popularity}{pop. variability}{pop. entropy}{pop. trend}
      & \cuecell{last-item popularity}{suffix pop. variability}{suffix pop. entropy}{pop. shift vs. mid} \\
    \bottomrule
  \end{tabular}
\end{table*}

Cue scalars summarize level, variability, trend, switching, or repetition. Macro uses the five most recent completed sessions, mid the observed active-session prefix, and micro the recent local suffix. Together, these scopes expose cross-session tendencies, behavior unfolding in the active session, and immediate local changes without using the prediction target.

\section{Experimental Details}
\label{app:data-details}

This appendix reports data processing, model selection, and computational efficiency. All experiments use one NVIDIA RTX 4080 SUPER GPU (16\,GB) per run.

\begin{figure}[!t]
  \centering
  \NewFig[width=\columnwidth]{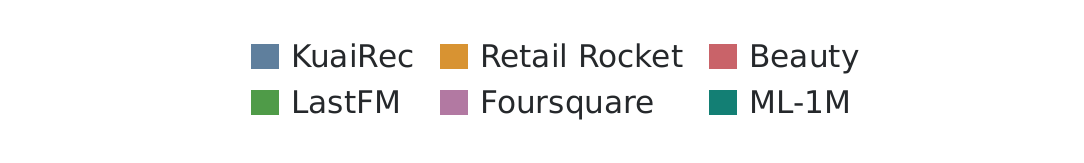}
  \vspace{0.3em}\\
  \NewFig[width=\columnwidth]{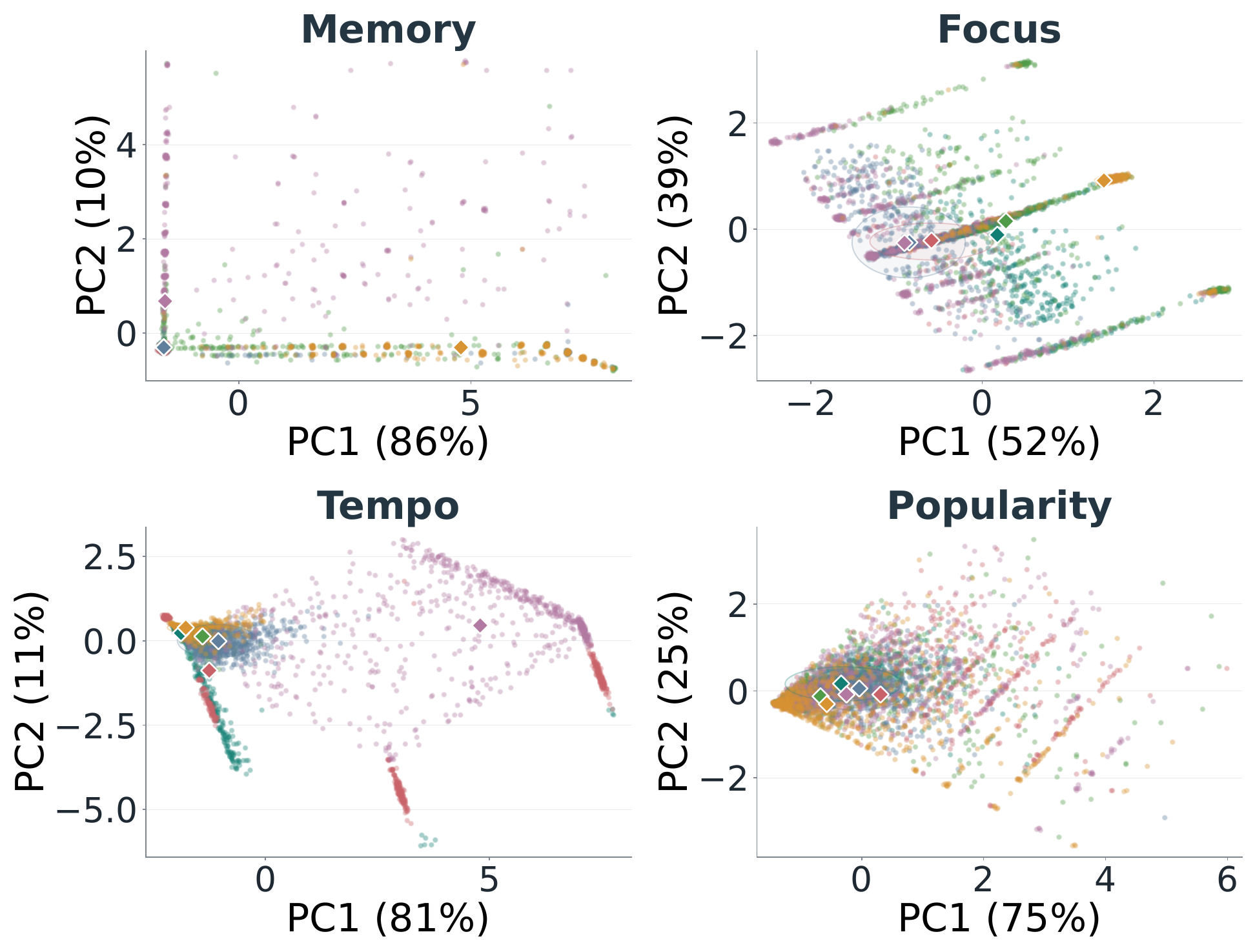}
  \caption{Behavioral cue diversity across six datasets, projected onto the first two principal components ($2{\times}2$ grid, one panel per family). Diamonds mark per-dataset medians. This extends Figure~\ref{fig:dataset-stat-evidence} to all four cue families.}
  \Description{Principal-component scatter plots for four cue families.}
  \label{fig:dataset-cue-diversity}
\end{figure}

\begin{table}[t]
  \caption{Processed dataset statistics used in the main experiments.}
  \label{tab:appendix-datasets}
  \centering
  \footnotesize
  \setlength{\tabcolsep}{3.6pt}
  \renewcommand{\arraystretch}{1.06}
  \begin{tabular}{@{}lp{0.15\columnwidth}cccc@{}}
    \toprule
    Dataset & Domain & \shortstack{Inter.\\(M)} & \shortstack{Sess.\\(K)} & \shortstack{Items\\(K)} & \shortstack{Avg.\\len.} \\
    \midrule
    KuaiRec & Video & 3.44 & 183.3 & 6.5 & 18.8 \\
    LastFM & Music & 14.23 & 658.6 & 426.2 & 21.6 \\
    Retail Rocket & Browse & 0.37 & 41.8 & 26.6 & 8.7 \\
    ML-1M & Movie & 0.57 & 14.2 & 3.1 & 40.4 \\
    Foursquare & POI & 0.05 & 4.6 & 3.6 & 9.9 \\
    Beauty & E-com. & 0.02 & 2.5 & 1.9 & 7.5 \\
    \bottomrule
  \end{tabular}
\end{table}
\subsection{Datasets and Preprocessing}

The datasets span short e-commerce sessions, location visits, ratings, videos, and long listening histories. Table~\ref{tab:appendix-datasets} reports their processed statistics, and Appendix~\ref{app:dataset-cue-diversity} visualizes cue-space diversity.

Data leakage and split strategy can distort offline recommender evaluation~\cite{ji2023leakage,gusak2025timesplit}. We preserve source session identifiers for Beauty, Foursquare, ML-1M, and Retail Rocket. KuaiRec and LastFM instead construct sessions using a 30-minute inactivity threshold. Sessions shorter than five interactions and items with fewer than three occurrences are removed. Sessions are split chronologically into 70/15/15 train/validation/test partitions; sessions longer than 50 interactions are then chunked within each partition, so chunks from one original session never cross splits. After fitting the training vocabulary, we remove validation/test sessions with training-unseen targets and filter training-unseen context items, giving every method the same closed candidate catalog. RouteRec cues use only observed history and training-split statistics. Frozen manifests and exact commands are provided in the public repository.

\subsection{Behavioral Cue Diversity}
\label{app:dataset-cue-diversity}

Figure~\ref{fig:dataset-cue-diversity} projects each family's session-level cue vectors onto two principal components (interquartile-range standardized and log-transformed where appropriate; 750 sessions per dataset). Cue structure varies by both dataset and family, with especially pronounced separation for KuaiRec and LastFM, and is not explained by average session length alone.
This projection is descriptive evidence of cue-space heterogeneity; it is not used to train the router or define dataset labels.

\subsection{Model Selection and Tuning}

All models are tuned within the search spaces in Table~\ref{tab:appendix-search-space}. We select the configuration maximizing the mean of validation HR@10, NDCG@10, and MRR@10, evaluate the test set only after selection, and report mean and sample standard deviation over seeds 42--44. Shared parameters apply to all models; in model-specific rows, ``Weight'' denotes the corresponding loss or feature-fusion coefficient.

\begin{table}[t]
  \centering
  \caption{Same-GPU efficiency results. Exec./logical is the ratio of actually executed to logically active non-item parameters; peak memory reports training/inference in MiB.}
  \label{tab:appendix-efficiency}
  \scriptsize
  \setlength{\tabcolsep}{3pt}
  \renewcommand{\arraystretch}{1.06}
  \begin{tabular}{@{}llrrrr@{}}
    \toprule
    Dataset & Model &
    \shortstack{Exec./\\logical} &
    \shortstack{Train\\examples/s $\uparrow$} &
    \shortstack{Full-sort\\ms/batch $\downarrow$} &
    \shortstack{Peak memory\\train / infer $\downarrow$} \\
    \midrule
    Foursquare & SASRec-wide & 1.00 & 4,913.4 & 2.887 & 363.1 / 163.3 \\
    Foursquare & Hidden MoE  & 1.62 & 3,220.7 & 16.259 & 593.7 / 198.0 \\
    Foursquare & RouteRec    & 1.62 & 2,978.5 & 17.491 & 593.7 / 197.4 \\
    \midrule
    KuaiRec & SASRec-wide & 1.00 & 5,121.5 & 4.722 & 652.7 / 288.3 \\
    KuaiRec & Hidden MoE  & 1.98 & 2,975.6 & 19.150 & 1,318.5 / 356.9 \\
    KuaiRec & RouteRec    & 1.98 & 2,841.4 & 20.289 & 1,319.4 / 356.0 \\
    \bottomrule
  \end{tabular}
\end{table}
\begin{table}[t]
  \centering
  \caption{Model-selection search spaces.}
  \label{tab:appendix-search-space}
  \scriptsize
  \setlength{\tabcolsep}{3pt}
  \renewcommand{\arraystretch}{1.03}
  \begin{tabular}{@{}p{0.25\columnwidth}p{0.71\columnwidth}@{}}
    \toprule
    Parameter & Candidate sets \\
    \midrule
    \multicolumn{2}{@{}l}{\textit{Shared optimization}} \\
    Learning rate & Log-uniform: KuaiRec $[3{\times}10^{-4},5{\times}10^{-3}]$; Foursquare/Retail Rocket/ML-1M $[1.5{\times}10^{-4},2.2{\times}10^{-3}]$; LastFM $[8{\times}10^{-5},1.2{\times}10^{-3}]$; Beauty $[1.5{\times}10^{-4},2{\times}10^{-3}]$ \\
    Weight decay & $\{5{\times}10^{-7},10^{-6},10^{-5},5{\times}10^{-5},10^{-4},1.5{\times}10^{-4}\}$ \\
    History length & $\{10,20,30\}$ or $\{10,20,30,50\}$ \\
    Hidden / inner width & $\{64,96,112,128,160\}$; inner $\{128,192,224,256,320\}$ (or $+32$ each) \\
    Layers / heads & $\{1,2,3,4\}$; $\{1,2,4,8\}$ \\
    Dropout & Hidden $\{0.10\text{--}0.18\}$; recurrent $\{0.10\text{--}0.30\}$; attention $\{0.06\text{--}0.20\}$ \\
    \midrule
    \multicolumn{2}{@{}l}{\textit{Baseline additions}} \\
    TiSASRec & Time span $\{64,128,256,384,512\}$ \\
    DuoRec/FEARec & Temperature $\{0.16\text{--}0.24\}$; contrastive weight $\{0.02\text{--}0.06\}$; semantic weight $\{0\text{--}0.12\}$ \\
    BSARec & $\alpha\{0.35,0.50,0.55,0.70\}$; $c\{2,3,5,7\}$ \\
    DIF-SR/FDSA & Attribute width $\{96,128,160,192\}$; attribute-fusion weight $\{0.08\text{--}0.15\}$; fusion $\{\text{gate,sum,concat}\}$ \\
    FAME & Expert count $\{2,3,4,5,6\}$ \\
    \midrule
    \multicolumn{2}{@{}l}{\textit{RouteRec additions}} \\
    Depth / dropout & Depth $\{1,2,3\}$; dropout $\{0.10\text{--}0.24\}$ \\
    Experts per group $K$ & $\{3,4,5,6\}$ \\
    Router width & $\{32,64,96,128\}$ \\
    Routing dropout & Cue $\{0,0.03,0.05,0.10\}$; attention $\{0.05\text{--}0.12\}$ \\
    Regularization & $\lambda_{\mathrm{cons}}\{0,2.5,5,8,12\}{\times}10^{-4}$; $\lambda_z\{0,5,10,20\}{\times}10^{-5}$ \\
    \bottomrule
  \end{tabular}
\end{table}

\newpage
\subsection{Computational Efficiency}
\label{app:efficiency}

We benchmark practical efficiency on a single NVIDIA RTX 4080 SUPER, running all models sequentially with the same data split, sequence length, candidate set, and batch size (256). Training throughput is measured over one representative epoch, while full-sort latency is measured on the same fixed validation batch after 20 warm-up batches and 100 timed batches, with three independent repetitions. SASRec-wide matches RouteRec's logical active non-item parameter count within 0.4\%, and Hidden MoE uses the same expert structure and dense execution as RouteRec but replaces behavior-guided routing with hidden-state routing.

Compared with SASRec-wide, RouteRec retains 60.6\% and 55.5\% of training throughput and incurs $6.06\times$ and $4.30\times$ full-sort latency on Foursquare and KuaiRec, respectively. Against the execution-matched Hidden MoE, however, RouteRec retains 92.5\% and 95.5\% of training throughput.
Because Hidden MoE and RouteRec share the same dense expert execution, their similar runtime isolates the modest additional cost of explicit cue conditioning. The benchmark supports a routing-quality contribution, not a realized sparse-efficiency advantage.

\clearpage
\section*{GenAI Usage Disclosure}
{\small
The authors used generative AI tools for language editing, translation, grammar and manuscript checking, limited assistance with implementing and debugging routine code under author-specified experimental settings, and organizing existing outputs for verification and plot-formatting checks. The tools did not determine research ideas, experimental design, parameter choices, substantive analyses, result interpretation, or conclusions, and did not alter underlying experimental values. All AI-assisted outputs were reviewed and verified by the authors, who take full responsibility for the final content.
\par}

\bibliographystyle{ACM-Reference-Format}
\bibliography{refs}

\end{document}